\documentclass[pra,twocolumn,preprintnumbers,amsmath,amssymb]{revtex4}

\usepackage{graphicx}% Include figure files
\usepackage{epstopdf}
\usepackage{amscd}

\usepackage{dcolumn}% Align table columns on decimal point
\usepackage{bm}% bold math

\newcommand{\proj}{\operatorname{proj}}  % define "proj" as a command

\newcommand{\vect}[1] {\ensuremath{\mathbf{#1}}} % display of N dim real vectors, or 3D vectors,
\newcommand{\set}[1] {\ensuremath{\mathcal{#1}}} % display of SETS like the measurement set
\newcommand{\mment}[1] {\ensuremath{\mathbf{#1}}} % display of MEASUREMENTS (i.e. devices with particular settings)
\newcommand{\cvect}[1] {\textsf{#1}} % display of N dim complex vectors

\newcommand{\cmatrix}[1]{\textsf{#1}} % Display of complex matrices

\newcommand{\rmatrix}[1]{\ensuremath{#1}} % Display of real matrices

\newcommand{\model}[1]{\ensuremath{\mathbf{#1}}} % display of q( ) symbol

\newcommand{\ev}[1]{\ensuremath{\langle #1 \rangle}} % Display of expectation values
\newcommand{\evb}[1]{\ensuremath{\bigl\langle #1 \bigr\rangle}} % Display of expectation values

\newcommand{\var}{\ensuremath{\chi}} % Symbol for what currently called \phi
\addtocounter{secnumdepth}{1} %Number down to the paragraph level

\begin{document}

%\setdefaultleftmargin{0em}{}{}{}{}{}

\title{An Information-Geometric Reconstruction of Quantum Theory,~II: \\
        The Correspondence Rules of Quantum Theory}

\author{Philip Goyal}
    \email{pgoyal@perimeterinstitute.ca}
    \affiliation{Perimeter Institute \\ Waterloo, Canada}


\begin{abstract}
In a companion paper~(hereafter referred to as
Paper~I), we have presented an attempt to derive the
finite-dimensional abstract quantum formalism within the framework of information geometry.
In this paper, we
formulate a correspondence principle, the \emph{Average-Value
Correspondence Principle}, that allows relations between
measurement results which are known to hold in a classical model
of a system to be systematically taken over into the quantum model
of the system. Using this principle, we derive the explicit form
of the temporal evolution operator~(thereby completing the
derivation of the abstract quantum formalism begun in Paper~I),
and derive many of the correspondence rules~(such as operator
rules, commutation relations, and Dirac's Poisson bracket rule)
that are needed to apply the abstract quantum formalism to model
particular physical systems.

\end{abstract}

         % Classification Scheme.
\maketitle

\section{Introduction}

In order to obtain a quantum mechanical model for a particular
physical system such as a particle moving in space, it is necessary
to supplement the abstract quantum formalism with various rules, to which we shall refer
collectively as \emph{correspondence rules},
which explicitly determine the form of the operators that represent
particular measurements performed on the system, or that represent
particular symmetry transformations~(such as displacement or
rotation) of the frame of reference in which the system is being
observed.  These correspondence rules can be usefully
classified as follows:
\begin{itemize}
\item[(i)]\emph{Operator Rules.} Rules for writing down operators
representing measurements of observables that, in the framework of
classical physics, are known functions of other, elementary,
observables whose operators are given~%
\footnote{The basic operator rules of quantum theory
are~(i)~Rule~1~(Function rule):~If~$A \rightarrow\cmatrix{A}$, then~$f(A)\rightarrow f(\cmatrix{A})$; (ii)~Rule~2~(Sum rule):~If~$A \rightarrow\cmatrix{A}$ and~$B\rightarrow\cmatrix{B}$, then~$f_1(A) + f_2(B) \rightarrow f_1(\cmatrix{A})
+ f_2(\cmatrix{B})$~(and similarly for more than two observables),
and (iii)~Rule~3~(Product rule):~If~$A \rightarrow\cmatrix{A}$ and~$B\rightarrow\cmatrix{B}$,
where~$[\cmatrix{A}, \cmatrix{B}]=0$, then~$f_1(A)f_2(B) \rightarrow f_1(\cmatrix{A})f_2(\cmatrix{B})$~(and similarly for more than two observables).  A more general rule that is often employed
is:~(iv)~Rule~3$'$~(Hermitization rule):~If~$A \rightarrow\cmatrix{A}$ and~$B\rightarrow\cmatrix{B}$,
then~$f_1(A)f_2(B) \rightarrow (f_1(\cmatrix{A})f_2(\cmatrix{B}) +
f_2(\cmatrix{B})f_1(\cmatrix{A}))/2$.}.
For example, such rules are needed to be able to write down the
operator that represents a measurement of~$H$, given the classical
relation~$H = p_x^2/2m + V(x)$, in terms of the
operators~$\cmatrix{x}$ and~$\cmatrix{p}_x$ which represent
measurements of~$x$ and~$p_x$, respectively.

\item[(ii)]\emph{Measurement Commutation Relations.}  Commutation relations
between operators that represent measurements of fundamental
observables such as position, momentum, and components of angular
momentum.  The commutation
relations~$[\cmatrix{x},\cmatrix{p}_x]=i\hbar$ and~$[\cmatrix{L}_x,
\cmatrix{L}_y]=i\hbar\cmatrix{L}_z$, are the obvious examples, while
Dirac's Poisson bracket rule,~$[\cmatrix{A}, \cmatrix{B}] = i\hbar
\widehat{\{A, B\}}$, is the more general rule, where~$\{A,B\}$ is the classical Poisson
bracket for observables~$A$ and~$B$,
and~$\widehat{\{A,B\}}$, $\cmatrix{A}$, and~$\cmatrix{B}$ are the
respective operators.

\item[(iii)]\emph{Measurement--Transformation Commutation Relations.} The
commutation relations between measurement operators and transformation
operators. For example, in the case of the $x$-displacement
operator,~$\cmatrix{D}_x$, one has, for example,~$[\cmatrix{x},\cmatrix{D}_x]=i$ and~$[\cmatrix{p}_x, \cmatrix{D}_x]=0$.  

\item[(iv)]\emph{Transformation Operators.}  Explicit forms of
the operators that represent symmetry transformations of a frame
of reference, such as the $x$-displacement operator~$\cmatrix{D}_x
= -i\, d/dx$.

\end{itemize}

The physical origin of many of the above-mentioned rules is obscure.
For example, although one can give simple physical
arguments~\footnote{See~\cite{Dirac58}~(Sec.~11)
and~\cite{vNeumann55}~(Sec.~IV.1), for example.} for the operator
rule which states that, if a measurement of~$A$ is represented by
operator~$\cmatrix{A}$, then a measurement of the function~$f(A)$
is represented by operator~$f(\cmatrix{A})$, the generalization of
such arguments to measurements of functions of two or more
observables encounters severe difficulties due to the possible
non-commutativity of the operators that represent these observables.
As a result of such difficulties, operator rules tend to be
heuristic and tend to lead to inconsistencies when applied to
particular examples~\footnote{See, for
example,~\cite{vNeumann55}~(Sec.~IV.1),
\cite{Bohm51}~(Sec.~9.12---9.15), and~\cite{Isham-QT}~(Sec.~5.2.1).
We shall discuss one such example in
Sec.~\ref{sec:inconsistencies}.}. Similarly, the commutation
relationship~$[\cmatrix{x},\cmatrix{p}_x]=i\hbar$ is typically
obtained from Schroedinger's equation or from Dirac's Poisson
bracket rule, both of whose derivations involve abstract assumptions
whose physical origin is obscure.

The question of what additional physical ideas are needed to obtain the
correspondence rules described above \emph{given} the abstract
quantum formalism has received relatively little recent attention.
Recent work on the elucidation of the physical origin of the
quantum formalism~(see Paper~I for references~\cite{Goyal-QT1b}) tends to neglect the question of the physical origin of the 
correspondence rules or  is concerned with the derivation of the
Schroedinger equation directly from physical ideas without taking
the abstract quantum formalism as a
given~\cite{Frieden-Schroedinger-derivation,
MJWHall-Schroedinger-derivation,
Reginatto-Schroedinger-derivation}. 

Operator rules have been discussed in a few publications, for
example in~\cite{vNeumann55, Groenewold-Principles-QM}, but a
derivation of the operator rules on the basis of a physical
principle, taking the abstract formalism as a given, has not been
successfully completed. The most recent systematic attempt to derive
the commonly employed measurement and
measurement--transformation commutation relations in such a manner is found
in~\cite{TFJordan75}~(see also Refs.~\cite{TFJordan69,
Ballentine98}). In~\cite{TFJordan75}, using the commutation relationships
for the operators that represent the Galilean group of
transformations, and establishing
the relation between these transformation operators and particular
measurement operators, commutation relations for these measurement
operators are obtained. However, this approach implicitly makes use
of operator rules, and relies upon auxiliary assumptions~(such as
the assumption that certain measurement operators are unchanged by
the action of particular symmetry operators) whose physical origin
is unclear.

In this paper, we show that, starting from the abstract quantum
formalism, it is possible to derive all the above-mentioned
correspondence rules in a systematic manner from a single
correspondence principle.   The principle, called the Average-value Correspondence Principle~(AVCP),
rests on the simple, intuitive notion that the quantum and classical models of the same physical system~(such as a particle moving in a potential) should agree in their predictions if one compares the expected values of the quantum measurements with the values obtained from the corresponding classical measurements, a connection which is known, via Ehrenfest's theorem, to hold approximately between the classical and quantum models of a particle~\cite{Ehrenfest-average-values}.

The principle expresses the classical-quantum correspondence in the following form:~in a classical experiment, if a measurement of observable~$A$ on a physical system can be implemented by an arrangement where measurements of  observables~$A', A'',\dots$ are performed on the system~(possibly performed at a different time to the measurement of~$A$), such that the value obtained from measurement of~$A$ can be calculated as a function,~$f$, of the values obtained from the measurements of~$A', A'',\dots$, then we can construct a corresponding quantum experiment, where quantum measurements~$\mment{A}, \mment{A}', \mment{A}'',\dots$, corresponding to classical measurements of~$A, A', A'',\dots$, are performed on copies of the same physical system, such that the same functional relation holds \emph{on average} between the values obtained from the quantum measurements, where the average is taken over infinitely many trials of the quantum experiment.  This connection holds only if~$f$ satisfies a particular condition.  The particular form of the quantum experiment is stipulated by the AVCP, and consists in specifying whether, for each pair of quantum measurements, the measurements in the pair are performed on the same or different copies of the system.  

This paper is organized as follows. We begin in Sec.~\ref{sec:AVCP}
by formulating the~AVCP. In Sec.~\ref{sec:deduction}, the AVCP is
used to obtain several \emph{generalized operator rules} which connect the
average values of operators at different times, from which the
commonly used operator rules of quantum theory follow as a special
case. Using the AVCP and the Temporal Evolution postulate~(from Paper~I), we then
derive the explicit form of the temporal evolution operator, thereby completing
the derivation of the finite-dimensional abstract quantum
formalism begun in Paper~I.

Next, in Sec.~\ref{sec:commutation-relations}, taking the
infinite-dimensional form of the abstract quantum formalism as a
given, we use the AVCP to derive many of the commonly employed
correspondence rules of quantum theory, namely~(a) the commutation
relations~$[\cmatrix{x},\cmatrix{p}_x]=i\hbar$
and~$[\cmatrix{L}_x, \cmatrix{L}_y]=i\hbar\cmatrix{L}_z$, and
Dirac's Poisson bracket rule, (b)~the explicit form of the
operators for displacements and rotations, and (c)~the commutation relations
between momentum and displacement operators, and between angular
momentum and rotation operators.

We note that the treatment of the correspondence rules is
illustrative rather than exhaustive, so that only the most
commonly encountered measurement and transformation operators
which are needed to formulate non-relativistic and relativistic
quantum mechanics have been discussed. Many other correspondence
rules~(such as the operators that represent Galilei
transformations, temporal displacement, and discrete
transformations such as spatial inversion) can be obtained by
arguments which closely follow those presented. The paper
concludes with a discussion of the results obtained.

\section{The Average-Value Correspondence Principle}
\label{sec:AVCP}

\subsection{Introduction}
\label{sec:AVCP-introduction}

Suppose that, as described in Paper~I, a quantum
model~$\model{q}(N)$, of dimension~$N$, has been constructed to
describe an abstract experimental set-up consisting of a source of
identical systems, a measurement set~$\set{A}$, and an interaction
set~$\set{I}$.  The measurements in~$\set{A}$, and the degenerate
forms of measurements in~$\set{A}$~\footnote{
A measurement that is a degenerate form of a
measurement~$\mment{A}$ is defined operationally in Paper~I.  Such
a measurement has~$N'<N$ possible results, and can be represented
as an $N$-dimensional degenerate Hermitian operator~(with~$N'$
distinct eigenvalues).},
%,
are represented by $N$-dimensional Hermitian operators~(possibly
with degenerate eigenvalues), and shall be called \emph{quantum}
measurements.

Suppose that quantum measurement~$\mment{A}$, with
operator~$\cmatrix{A}$, represents a measurement that is classically
described as a measurement of some observable~$A$~(which we shall
henceforth abbreviate to~``quantum measurement~$\mment{A}$~(or
operator~$\cmatrix{A}$) represents a (classical)~measurement of~$A$''). Suppose
that we wish to determine whether there is a quantum measurement
that represents a measurement of, say,~$A^2$, and, if so, to
determine the operator which can be said to represent this
measurement. Classically, one can imagine that a measurement
of~$A^2$ is implemented by an experimental arrangement where a measurement of~$A$ is
performed and the value obtained is then squared. If this arrangement is
described in the quantum model, it follows that, if the input state
is an element,~$\cvect{v}_i$, of an orthonormal set of eigenvectors
of~$\cmatrix{A}$, the output state of the process is~$\cvect{v}_i$
and the output of the process is~$a_i^2$,
where~$\cmatrix{A}\cvect{v}_i = a_i \cvect{v}_i$~($i=1,2, \dots,
N$).  According to this line of reasoning, it follows at once that
the operator~$\sum_i a_i^2\cvect{v}_i^\dagger\cvect{v}_i = \cmatrix{A}^2$ therefore represents a measurement
of~$A^2$.

However, this simple argument, given by Dirac~\cite{Dirac58}, does not readily generalize.  For
example, suppose that quantum measurements~$\mment{A}$
and~$\mment{B}$, with operators~$\cmatrix{A}$ and~$\cmatrix{B}$,
represent measurements of~$A$ and~$B$, respectively.  Suppose that
we wish to determine the operator~(if any exists) that represents a
measurement of~$A+B$. Classically, one imagines that such a
measurement is implemented by making measurements of~$A$ and~$B$
simultaneously on a system, and adding the values obtained. However,
if~$[\cmatrix{A}, \cmatrix{B}] \ne 0$, this arrangement cannot be
described in the quantum framework since, in this framework, the
measurements cannot be performed at the same time on a single system.  If one, then, instead considers an
arrangement where, for instance, a measurement of~$A$ is followed immediately by a measurement of~$B$, then, if the arrangement is described in the quantum framework, one cannot find~$N$ input states which emerge as output states since the non-commutativity of~$\cmatrix{A}$ and~$\cmatrix{B}$ implies that~$\cmatrix{A}$ and~$\cmatrix{B}$ share fewer than~$N$ eigenstates in common, making it impossible to construct a simple parallel to Dirac's argument.

Now, the conventional operator rules of quantum theory assert that a
measurement of~$A+B$ is represented by the
operator~$\cmatrix{A}+\cmatrix{B}$. Although this seems entirely
reasonable on a formal, symbolic level, it is not clear in what
physical sense~$\cmatrix{A}+\cmatrix{B}$ can be said to `represent'
the measurement.  To see this, note that, whereas the eigenvectors and eigenvalues
of~$\cmatrix{A}$ directly reflect the output states and
the values obtained when measurement~$\mment{A}$ is performed on the
system, the eigenvectors of~$\cmatrix{A}+\cmatrix{B}$ do not, in
general, coincide with the eigenvectors of either~$\cmatrix{A}$
or~$\cmatrix{B}$, and the eigenvalues of~$\cmatrix{A}+\cmatrix{B}$
do not, in general, coincide with the outputs of any
plausible arrangement~(described in the quantum model) that classically implements a
measurement of~$A+B$ which involves performing measurements of~$A$
and~$B$ and combining the values obtained. For example, a classical measurement of~$S_x + S_z$ is implemented by an arrangement where measurements of~$S_x$ and~$S_z$ are performed immediately after one another, and their values added.  When modeled in the quantum framework, the possible outputs of this arrangement, in the case where the measurements are performed on a spin-1/2 particle, are~$\pm \hbar, 0$, whereas a quantum measurement represented by operator~$\cmatrix{S}_x + \cmatrix{S}_z$  has possible values~$\pm \hbar/\sqrt{2}$.

The above observations illustrate the difficulty of obtaining a
physical understanding of the operator rules of quantum theory even
in simple cases of interest.   Below, we shall formulate a physical
principle which gives a clear physical meaning to the sense in which
an operator can be said to represent a classical
measurement, and, in the majority of cases of interest, uniquely
determines the operator which represents such a measurement.

\subsubsection{Implementations of classical
measurements}

\begin{figure}[!h]
\begin{centering}
\includegraphics[width=3.25in]{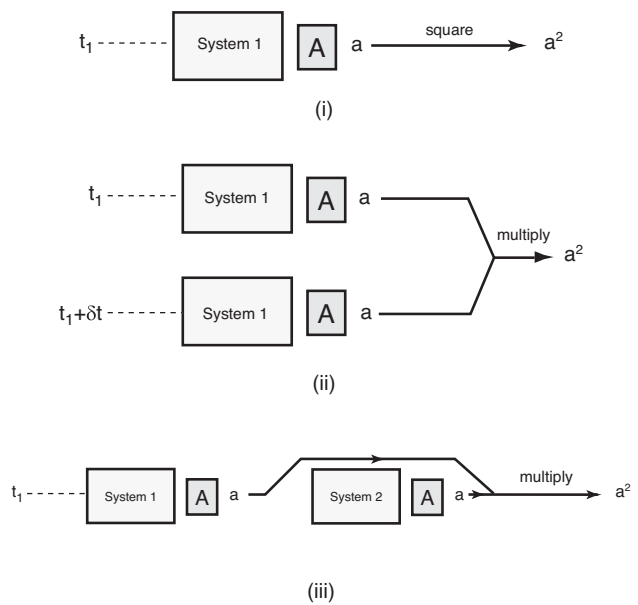}
\caption{\label{fig:mments-of-A-squared} Three arrangement that classically implement 
a measurement of~$A^2$.
In~(i), a measurement of~$A$ is made on one copy of the system, and
the value obtained is squared to give the output; in (ii), two immediately
successive measurements of~$A$ are made on one copy of the system,
and the output is obtained by multiplying the two values obtained; and
(iii)~two simultaneous measurements of~$A$ are made on two separate
copies of the system prepared in the same state, and the output is
obtained by multiplying the two values obtained. In a classical model of
these arrangements, each arrangement yields the same output. However,
in the quantum model of these arrangements, the expected output
of~(i) and~(ii) is~$\overline{a^2}$ whereas the expected output
of~(iii) is~$(\overline{a})^2$.}
\end{centering}
\end{figure}

Consider again a measurement of~$A^2$, where~$A$ is some observable.
Classically, one can imagine a measurement of~$A^2$ being
implemented using one of three different arrangements~(see
Fig.~\ref{fig:mments-of-A-squared}): (i)~make a measurement of~$A$
on one copy of the system, and square the value obtained; (ii)~make two
immediately successive measurements of~$A$ on one copy of the
system, and multiply the two values obtained; or~(iii)~make two
simultaneous measurements of~$A$ on two separate copies of the
system prepared in the same state, and multiply the two values obtained.
Although the first of these arrangements is the one we have
considered above, all three arrangements yield the same output
when modeled classically, and so can be regarded as equally valid
implementations of a measurement of~$A^2$ in the classical framework.

Now, perhaps surprisingly, when described using the quantum model,
these arrangements do not, in general, yield the same expected
outputs. Consider a quantum theoretic description of the first arrangement where quantum 
measurement~$\mment{A}$ represents a measurement of~$A$.  
In each run of the experiment, one copy of the
system is prepared in some given state. Let the probability that a
measurement~$\mment{A}$ yields result~$i$~($i=1, 2, \dots, N$), with
value~$a_i$, be denoted~$p_i$.  Then, the expected output is given
by
\begin{equation}
\begin{split}
\label{eqn:expected-outcome-example-1-implementation-1}
\text{Expected output} &= \sum_i (a_i)^2 p_i \\
                                &= \overline{a^2}
\end{split}
\end{equation}
One finds that arrangement~(ii) yields the same expected
output. However, in arrangement~(iii), where, in each run of
the experiment, two copies of the system are prepared in the same
state, one obtains
\begin{equation}
\label{eqn:expected-outcome-example-1-implementation-3}
\begin{split}
\text{Expected output} &= \sum_{i,j} (a_i a_j) p_ip'_j \\
                                &= (\overline{a})^2,
\end{split}
\end{equation}
with~$p_i$ and~$p'_j$ denoting respectively the probabilities that
measurements~$\mment{A}$ on the first and second copy yield
result~$i$ and~$j$~($i, j=1, 2, \dots, N$).

In this example, the differences between the arrangements as
viewed in the quantum model arise due to the fact that, in the
quantum framework, an immediate repetition of a measurement on the
same copy of a system is different from performing a second
simultaneous measurement on a separate, identically-prepared, copy
of the system.

\begin{figure}[!h]
\begin{centering}
\includegraphics[width=3.25in]{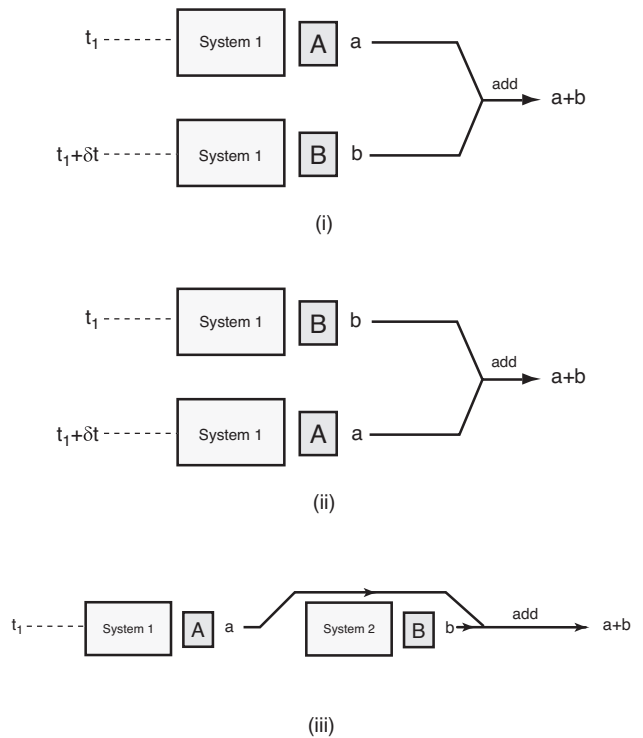}
\caption{\label{fig:mments-of-A-plus-B} Three implementations of the
measurement classically described as ``a measurement of~$A+B$''.
In~(i), a measurement of~$A$, and then a measurement of~$B$, is made
on one copy of the system, and the values obtained are then added to give
the output; in (ii), a measurement of~$B$, and then a measurement
of~$A$, is made on one copy of the system, and the values obtained are then
added to give the output; and (iii)~simultaneous measurements of~$A$
and~$B$ are made on two separate copies of the system prepared in
the same state, and the output is obtained by adding the two
values obtained. In a classical model of these arrangements, each
arrangement yields the same output. However, in the quantum model
of this situation, the three arrangements do not, in general,
yield the same expected outputs.}
\end{centering}
\end{figure}

In the case of a measurement of~$A+B$, where measurements of~$A$ and~$B$ are assumed
to occur at the same time, one needs to take into account the
additional fact that, in the quantum framework, the order in which
two measurements is performed is of possible importance.
Accordingly, one can imagine implementing a measurement of~$A+B$ using
one of at least three arrangements~(see
Fig.~\ref{fig:mments-of-A-plus-B}): (i)~make a measurement of~$A$,
and then a measurement of~$B$, on one copy of the system, and add
the values obtained; (ii) make a measurement of~$B$, and then a measurement
of~$A$, on one copy of the system, and add the values obtained; (iii)~make
a measurement of~$A$ on one copy of the system and, simultaneously,
a measurement of~$B$ on a second copy of the system prepared in the
same state as the first copy, and add the values obtained. Once again, in a
classical model of these arrangements, the outputs agree. However, if
one calculates the expected outputs in the quantum model, one finds
that, if~$[\cmatrix{A}, \cmatrix{B}]\ne 0$, all three will, in
general, disagree with one another.

As these examples illustrate, different arrangements that implement the same
classical measurement do not, in general, give the same
expected output when described in the quantum framework. 

\subsubsection{The average-value condition}
\label{sec:av-condition}

The existence of different arrangements that implement the same
classical measurement immediately raises two questions.
First, are these arrangements, in some sense, equally valid in
the quantum framework, or it is possible to find some reasonable
physical basis upon which to select particular arrangements and
regard these as more fundamental than the others? Second, is it
possible to find operators that represent the selected
arrangements, and, if so, do all the selected arrangements that implement
a given measurement have the same operator representation?

To answer these questions, we begin by observing that, although the
above arrangements are all regarded as \emph{bone fide} implementations of
measurements in the classical model, a measurement is only
describable as such in the quantum model, and so can be called a
quantum measurement, if it can be represented by a Hermitian
operator which represents a single measurement performed upon one
copy of the system at a particular time. So, for example, although
arrangement~(iii) of a measurement of~$A^2$ can be modeled in the
quantum framework, it cannot be described
as a quantum measurement since it involves two separate
measurements. In contrast, arrangement~(i) 
can be described as a quantum measurement since it only
involves a single measurement on one copy of the system.

However, although an arrangement 
involving more than one measurement cannot itself be regarded as
a quantum measurement when modeled in the quantum framework, 
we can reasonably ask whether it is possible
to find a quantum measurement,~$\mment{C}$, with
operator~$\cmatrix{C}$, which, in some sense to be determined, can
nonetheless be said to \emph{represent} the arrangement.

At the outset, we note that it makes no sense to require that
measurement~$\mment{C}$ always yield a value that coincides with the output of a given
arrangement since measurement results are only probabilistically
determined in the quantum framework. However, we \emph{can} impose
the simple condition that, over an infinite number of runs,  the
average value obtained from measurement~$\mment{C}$ should coincide with the
average output obtained from the arrangement for any initial
state of the system.

For example, in the case of an arrangement that implements a measurement
of~$A^2$, quantum measurement~\mment{C} which, by hypothesis, represents the
arrangement, must be such that, for all states of the
system,~$\ev{\cmatrix{C}}$ is equal to the expected output obtained
from the arrangement.  In the case of arrangements~(i)
and~(ii) described above, using
Eq.~\eqref{eqn:expected-outcome-example-1-implementation-1}, we
accordingly obtain the condition that the relation
\begin{equation}
\begin{split}
\ev{\cmatrix{C}}    &= \overline{a^2} \\
                    &= \ev{\cmatrix{A}^2}
\end{split}
\end{equation}
must hold for all states,~$\cvect{v}$, of the system.  From this
condition, we can conclude that
\begin{equation}
\label{eqn:avc-rule-1}
\cmatrix{C} = \cmatrix{A}^2.
\end{equation}
In the case of
arrangement~(iii), using
Eq.~\eqref{eqn:expected-outcome-example-1-implementation-3}, we
obtain the condition that the relation
\begin{equation}
%\label{}
\begin{split}
\ev{\cmatrix{C}}    &= (\overline{a})^2 \\
                    &= \ev{\cmatrix{A}}^2
\end{split}
\end{equation}
must hold for all~$\cvect{v}$.  By diagonalizing~$\cmatrix{A}$, one
can readily show that this condition implies that~$\cmatrix{A}$ is a
multiple of the identity, which represents a trivial measurement
that yields the same result irrespective of the state of the
system. Therefore, arrangement~(iii) does not satisfy the above
average-value condition in the case of any non-trivial measurement
of~$A$, and can therefore be reasonably eliminated. Hence, in this
case, the average-value condition is sufficiently strong so as to be
able to pick out arrangements~(i) and~(ii), and, since the
average-value condition also implies that these arrangements are
both represented by the operator,~$\cmatrix{A}^2$, one can
unambiguously conclude that a measurement of~$A^2$ is represented by
the operator~$\cmatrix{A}^2$.

Proceeding in a similar way, restricting ourselves for the time
being to measurements~$\mment{A}$ and~$\mment{B}$ that are not
subsystem measurements~\footnote{A subsystem measurement is
defined in Paper~I as a measurement performed on a single subsystem
of a composite system.}, one finds that, in the case of a
measurement of~$C = A+B$, only arrangement~(iii) is possible
if~$[\cmatrix{A}, \cmatrix{B}]\ne 0$, which then yields the operator
\begin{equation}
\label{eqn:avc-rule-2} \cmatrix{C} = \cmatrix{A}+\cmatrix{B}.
\end{equation}
If~$[\cmatrix{A}, \cmatrix{B}]= 0$, then all three arrangements
are possible, and all yield the same operator,~$\cmatrix{C}$, as
above.

Finally, in the case of a measurement of~$AB$, with~$[\cmatrix{A},
\cmatrix{B}]=0$, one finds that the arrangement must be the one
where measurements~\mment{A} and~\mment{B} are performed on the same
copy of the system, in which case the
operator~$\cmatrix{A}\cmatrix{B}$ is obtained.

Hence, we see that the average-value condition is sufficient to
yield a unique operator representation for the measurements
considered above.  Based on the above considerations, we can
tentatively formulate the following general rule:~in the case of a
measurement which is implemented by an arrangement that contains two elementary
measurements~(not subsystem measurements) represented by commuting
operators, it is possible to find a quantum measurement that
represents the arrangement if the two elementary measurements are
performed on the same copy of the system; but, when the operators do
not commute, the elementary measurements must be performed on
different copies of the system.

We now consider implementations of a measurement of~$AB$ in the case
when~$[\cmatrix{A}, \cmatrix{B}]\ne 0$. The general rule just given
suggests that we should consider the arrangement
where the measurements of~$A$ and~$B$ are performed on
different copies of the system.  When described in the quantum framework, 
this arrangement has the expected output
\begin{equation}
\sum_i \sum_j (a_i b_j) p_i p_j' =
\ev{\cmatrix{A}}\ev{\cmatrix{B}},
\end{equation}
with~$p_i$ and~$p'_j$ denoting respectively the probabilities that
the measurement~$\mment{A}$ on the first copy and the measurement~$\mment{B}$ on the second copy yield result~$i$ and~$j$~($i, j=1, 2,
\dots, N$), and~$a_i, b_j$ respectively denoting the values of the
$i$th and $j$th results of measurements~$\mment{A}$
and~$\mment{B}$. Imposing the above average-value condition, the
operator~$\cmatrix{C}$ that represents this implementation must
satisfy the condition
\begin{equation} \label{eqn:measurement-AB-av-desideratum}
\ev{\cmatrix{C}} =  \ev{\cmatrix{A}}\ev{\cmatrix{B}}
\end{equation}
for all~$\cvect{v}$.  However, for non-commuting~$\cmatrix{A}$
and~$\cmatrix{B}$, one finds that~$\cmatrix{C}$ cannot be found
such that this relation is satisfied for all~$\cvect{v}$. We note,
however, that with
\begin{equation}
\label{eqn:hermitized-form-AB} \cmatrix{C} = \frac{1}{2}
(\cmatrix{A}\cmatrix{B} + \cmatrix{B}\cmatrix{A}),
\end{equation}
equation~\eqref{eqn:measurement-AB-av-desideratum} holds for the
eigenstates of~$\cmatrix{A}$ and the eigenstates of~$\cmatrix{B}$,
which suggests that we may be able to weaken the average-value
condition in this case so that we only require that
Eq.~\eqref{eqn:measurement-AB-av-desideratum} holds for these
eigenstates.  However, as we shall illustrate
later~(Sec.~\ref{sec:inconsistencies}), the application of
Eq.~\eqref{eqn:hermitized-form-AB} can lead to inconsistencies.
Consequently, we conclude that, from the point of view of the
average-value condition, it is not possible to find an operator that
represents a measurement of~$AB$ when the operators~$\cmatrix{A}$
and~$\cmatrix{B}$ do not commute. More generally, we find that,
when~$[\cmatrix{A}, \cmatrix{B}]\ne 0$, it is necessary to exclude
measurements of~$f(A,B)$, with~$f$ analytic~(so that~$f$ has a
well-defined polynomial expansion in~$A$ and~$B$), where the
polynomial expansion of~$f$ contains product terms.

Finally, in the case of a classical measurement which is implemented by an 
arrangement which contains two measurements that are
performed on different subsystems of a composite system, one finds
that the arrangement can be represented by a quantum measurement
irrespective of whether the two measurements are performed on the
same or on different copies of the system, and that the different
possible arrangements are represented by the same operator.

\subsubsection{Generalizations}

In the examples above, we have considered arrangements that implement a
classical measurement in which the measurement and the
measurements in the arrangement are performed at the
same time, and in the same frame of reference. However, as the
following examples illustrate, these are unnecessary restrictions.

First, classically, for a non-relativistic particle of mass~$m$, one
can implement a measurement of~$x$ performed at time~$t+\delta t$ by
an arrangement where measurements of~$x$ and~$p_x$ are performed at
time~$t$, and the function~$x + p_x \,\delta t/m$ is then computed.

Second, one can implement a measurement of~$x'$ on a particle in the
reference frame~$S'$ that is displaced along the $x$-axis relative
to frame~$S$ by performing a measurement of~$x$ in frame~$S$, and
computing the appropriate function~$x'=f(x)$ that relates~$x$
and~$x'$.

The above considerations regarding the implementation of a
classical measurement are applicable without change to
the case where the measurement and the measurements in
the arrangement by which it is implemented are performed at different times or in
different frames of reference. Below, we shall accordingly
generalize our notion of the implementation of a
classical measurement.

\subsection{Statement of the AVCP}
\label{eqn:AVCP-statement}

\begin{figure}[!h]
\begin{centering}
\includegraphics[width=3.25in]{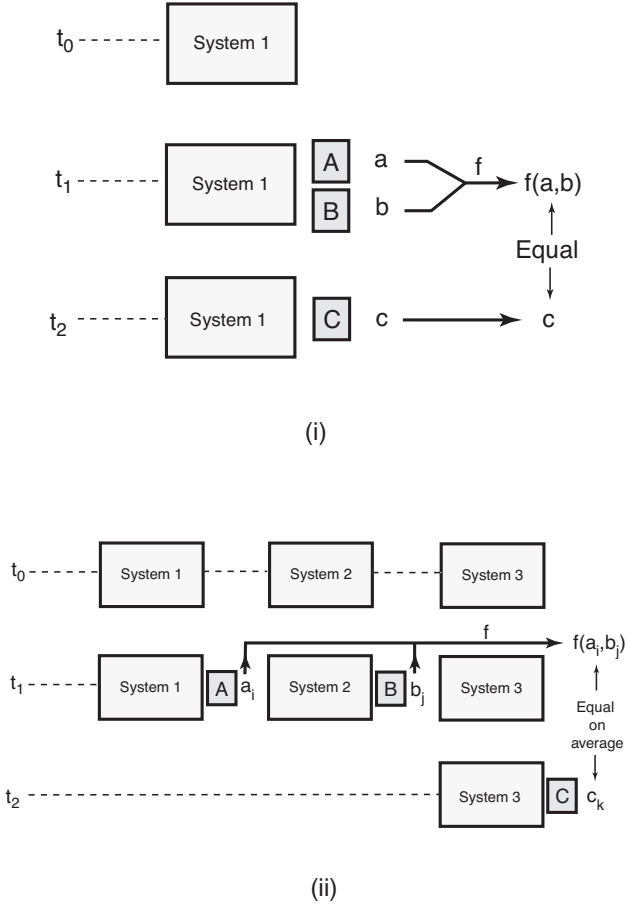}
\caption{\label{fig:AVCP-example} An example of the application of
the AVCP.  (i) A classical experiment showing the measurements
of~$A, B$ and~$C$ performed at times~$t_A, t_B$, and~$t_C$,
respectively, with values denoted as~$a, b$, and~$c$,
respectively. Here,~$t_A = t_B = t_1$ and~$t_C = t_2$. Suppose that
one finds that the relation~$c=f(a,b)$ holds for all initial states
of the system. (ii) The corresponding quantum experiment. Three
copies of the system are prepared in the same initial
state,~$\cvect{v}_0$, at time~$t_0$, and are placed in identical
backgrounds.  In this example, it is assumed that the
operators~$\cmatrix{A}$ and~$\cmatrix{B}$ do not commute. Hence, by
the AVCP, measurements~\mment{A} and~\mment{B} are performed on
different copies of the system. Measurement~\mment{C} is performed
on a separate copy of the system.  In any given run of the
experiment, the probabilities that measurements~\mment{A}, \mment{B}
and~\mment{C} yield values~$a_i, b_j$ and~$c_k$~($i, j, k =
1, \dots, N$), are~$p_i, p'_j$ and~$p_k''$, respectively. The AVCP
then asserts that, provided the polynomial expansion of~$f(a,b)$
contains no product terms involving~$a$ and~$b$, the
relation~$\overline{c}=\overline{f(a,b)}$ holds for all initial
states,~$\cvect{v}_0$, of the system, where the average is taken
over an infinite number of runs of the experiment; that is, $\sum_k
c_k p_k'' = \sum_{ij} f(a_i,b_j) p_i p'_j$ for all~$\cvect{v}_0$.}
\end{centering}
\end{figure}

We shall now state a general principle which incorporates the
observations made above concerning the average-value condition. An
illustrative example is given in Fig.~\ref{fig:AVCP-example}.

\begin{itemize}
   \item[]\textbf{Average-Value Correspondence Principle}
    Consider a classical idealized experiment in which a system~(possibly
    a composite system) is
    prepared in some state at time~$t_0$, and is allowed to evolve
    in a given background. Suppose that a
    measurement of~$A^{(m)}$~$(m\ge 2)$, performed on the
    system at time~$t_2$ with value~$a^{(m)}$, can be implemented
    by an arrangement where measurements of~$A^{(1)},
    A^{(2)}, \dots, A^{(m-1)}$ are performed upon one copy of the system
    at time~$t_1$, and the values of their respective results,
    denoted~$a^{(1)}, a^{(2)}, \dots,
    a^{(m-1)}$, are then used to compute the output~$f(a^{(1)},
    a^{(2)}, \dots, a^{(m-1)})$, where~$f$ is an analytic
    function, so that the relation
    \begin{equation*}
    \label{eqn:avc}
    a^{(m)}=f(a^{(1)}, a^{(2)}, \dots, a^{(m-1)}) \tag{$*$}
    \end{equation*}
    holds for all initial (classical)~states
    of the system.

    Consider the case where the quantum measurements~$\mment{A}^{(1)}, \mment{A}^{(2)}, \dots,
    \mment{A}^{(m)}$, with operators~$\cmatrix{A}^{(1)}, \cmatrix{A}^{(2)},
    \dots, \cmatrix{A}^{(m)}$, represent the
    measurements of~$A^{(1)}, A^{(2)}, \dots, A^{(m)}$, respectively.  Then, consider
    the following idealized quantum experimental
    arrangement consisting of several set-ups, each consisting of identical
    sources and backgrounds, where, in each set-up, a copy of the system is
    prepared in the same initial state,~$\cvect{v}_0$, at time~$t_0$.

    In one set-up, only measurement~$\mment{A}^{(m)}$ is
    performed~(at time~$t_2$) and, for any~$i,j$ with~$i\neq j$
    and~$i,j \leq m-1$, the measurements~$\mment{A}^{(i)},
    \mment{A}^{(j)}$ are performed~(at time~$t_1$) in:
    	\begin{enumerate}
	\item the \emph{same}
    set-up if both~$\bigl[\cmatrix{A}^{(i)},
    \cmatrix{A}^{(j)}\bigr]=0$ and, if the system is
    composite, the measurements are performed on the same
    subsystem; 
    	\item \emph{different} set-ups if~$\bigl[\cmatrix{A}^{(i)},
    \cmatrix{A}^{(j)}\bigr]\ne 0$.
	\end{enumerate}
	
    Let the values of the results of the measurements~$\mment{A}^{(1)}, \dots,
    \mment{A}^{(m)}$ in any given run of the experimental arrangement be
    denoted~$a^{(1)}, \dots, a^{(m)}$, respectively.
    The function~$f(a^{(1)}, a^{(2)}, \dots, a^{(m-1)})$ is defined
    as \emph{simple} provided that its
    polynomial expansion contains no terms involving a product of
    eigenvalues belonging to measurements whose operators
    do not commute.  If~$f$ is simple, then~\eqref{eqn:avc} holds
    on average, the average being taken over an infinitely large
    number of runs of the experiment.

\end{itemize}

The above principle can be generalized in a number of ways, for
example to the case where the measurements of~$A^{(1)}, \dots,
A^{(m-1)}$ are not performed at the same time. However, these
generalizations are unnecessary for the derivations of the usual
correspondence rules of quantum theory, and are therefore not
discussed here.

\section{Generalized Operator Rules and the Temporal Evolution Operator}
\label{sec:deduction}

\subsection{Generalized Operator Rules}

We will now apply the AVCP to derive operator relations which hold
when the function~$f$ takes various useful forms.  In each
instance of~$f$, we shall first derive a generalized operator rule
which relates the expected values of the relevant operators at
\emph{different times}.  Then, taking the special case when the
expected values are computed at the same time, we obtain the
corresponding operator rule which relates the operators
themselves.

We shall consider a classical experiment where a system is subject
to measurements of~$A$ and~$B$ at time~$t_1$, and to a measurement
of~$C$ at time~$t_2$. We shall suppose that a measurement of~$C$, with
value~$c$, can be implemented by an arrangement in which the
measurements of~$A$ and~$B$ are performed, with respective
values~$a$ and~$b$, and the function~$f(a,b)$ then computed, so
that the relation
\begin{equation}
c= f(a,b)
\end{equation}
holds for all initial states of the system.

In a quantum model of the appropriate experimental arrangement, let
the operators that represent these measurements be
denoted~$\cmatrix{A}$,~$\cmatrix{B}$, and~$\cmatrix{C}$,
respectively.  To simplify the presentation, we shall only consider
the case where these operators have finite dimension,~$N$; the
results obtained below can be readily shown to apply in the infinite
dimensional case. Let the elements of orthonormal sets of
eigenvectors of~$\cmatrix{A}, \cmatrix{B}$ and~$\cmatrix{C}$ be
denoted~$\cvect{v}_i, \cvect{v}_j'$, and~$\cvect{v}''_k$,
respectively~$(i, j, k = 1,2, \dots, N)$, let the corresponding
eigenvalues be denoted~$a_i, b_j$ and~$c_k$, and let the
probabilities of the~$i$th,~$j$th and~$k$th results of
measurements~$\mment{A}$,~$\mment{B}$ and~$\mment{C}$ in any given
experimental arrangement be denoted by~$p_i$,~$p'_j$ and~$p_k''$,
respectively.

\subsubsection*{Case 1. $f$ is a function of~$a$ only.}
\label{sec:AVCP-case1}

In this case, the quantum experiment simply consists of two
set-ups, involving two copies of the system,
where~$\mment{A}$ is performed on one copy at time~$t_1$
and~$\mment{C}$ on the other copy at time~$t_2$. Since
function~$f$ is simple, by the AVCP, the relation
\begin{equation} \label{eqn:AVCP-case-1-main-eqn}
    \sum_k c_k p_k'' =  \sum_i f(a_i) p_i
\end{equation}
holds for all initial states,~$\cvect{v}_0$, of the system.
Explicitly,
\begin{equation}
\begin{aligned}
p_i &= \big|\cvect{v}_i^\dagger \cvect{v}_{t_1}\big|^2 \\
p_k'' &= \big|\cvect{v}_k''^\dagger \cvect{v}_{t_2}\big|^2,
\end{aligned}
\end{equation}
with~$\cvect{v}_t$ being the state of the relevant copy of the
system at time~$t$.  Hence, we can write
Eq.~\eqref{eqn:AVCP-case-1-main-eqn} as
\begin{equation}
\cvect{v}^\dagger_{t_2}
                        \bigg(
                        \sum_k \cvect{v}_k'' {\cvect{v}''_k}^\dagger
                        c_k
                        \bigg)
\cvect{v}_{t_2} = \cvect{v}^\dagger_{t_1} \bigg(
                        \sum_i \cvect{v}_i \cvect{v}_i^\dagger f(a_i)
                        \bigg)
\cvect{v}_{t_1}.
\end{equation}
Noting that
\begin{equation}
\begin{aligned}
f(\cmatrix{A}) &= \sum_i \cvect{v}_i \cvect{v}_i^\dagger f(a_i) \\
\cmatrix{C} &= \sum_k \cvect{v}_k'' {\cvect{v}''_k}^\dagger c_k,
\end{aligned}
\end{equation}
we obtain the relation
\begin{equation} \label{eqn:AVCP-case-1-main-result}
\ev{\cmatrix{C}}_{t_2} = \evb{f(\cmatrix{A})}_{t_1},
\end{equation}
which holds for all~$\cvect{v}_0$.  We can summarize the above
result in the form of the \emph{generalized function rule}:
\begin{equation} \label{eqn:AVCP-case-1-general-rule}
c(t_2) = f\left(a(t_1)\right) \mapsto \ev{\cmatrix{C}}_{t_2} =
\evb{f(\cmatrix{A})}_{t_1} \quad\forall \cvect{v}_0,
\end{equation}
where, for clarity, the times at which the results are obtained has
been explicitly indicated.
In the special case where~$t=t_1 = t_2$, we obtain the usual
operator rule, the \emph{function rule}:
\begin{equation} \label{eqn:AVCP-case-1-special-case-result}
c = f(a) \mapsto  \cvect{C} =  f(\cvect{A}).
\end{equation}

\subsubsection*{Case 2. $f(a,b)= f_1(a) +  f_2(b)$}
\label{sec:AVCP-case2}

It is necessary to consider three sub-cases.  First, if
measurements~\mment{A} and~\mment{B} have commuting operators and,
in the case of a composite system, if they are subsystem
measurements performed on the same subsystem, then, by the AVCP,
they are performed on the same copy of the system in the quantum
experiment. Since~$f$ is simple, the AVCP applies, so that
\begin{equation}
\begin{split}
    \sum_k c_k p_k''    &=  \sum_i \bigg( f_1(a_i)  + \sum_j  f_2(b_j)
                                                             p'_{j|i}
                                    \bigg) p_i \\
                        &=   \sum_i \big( f_1(a_i)  +  f_2(b_i)
                                    \big) p_i,
\end{split}
\end{equation}
holds for all initial states,~$\cvect{v}_0$, of the system.  Here,
the notation~$p'_{j|i}$ is the probability that
measurement~\mment{B} yields result~$j$ given that~\mment{A} has
yielded result~$i$; in this case,~$p'_{j|i} = \delta_{ij}$.  From
the above relation, we obtain the generalized operator relation
\begin{equation} \label{eqn:AVCP-case2-av-relation}
\ev{\cmatrix{C}}_{t_2} = \evb{f_1(\cmatrix{A})}_{t_1} +
\evb{f_2(\cmatrix{B})}_{t_1},
\end{equation}
which holds for all initial states,~$\cvect{v}_0$.  In the special
case where~$t_1=t_2$, we obtain the operator relation,
\begin{equation} \label{eqn:C-operator1}
    \cmatrix{C} =  f_1(\cmatrix{A}) +  f_2(\cmatrix{B}).
\end{equation}

Second, in the case where measurements~\mment{A} and~\mment{B} are
subsystem measurements performed on different subsystems, they
can, by the AVCP, be performed on the same copy of the system, in
which case we obtain the same results as above.

Third, if measurements~\mment{A} and~\mment{B} have non-commuting
operators, then, by the AVCP, they are performed on different
copies of the system in the quantum  experiment. Since~$f$ is
simple, the AVCP again applies, so that the relation
\begin{equation}
    \sum_k c_k p_k''    = \sum_i   f_1(a_i) p_i + \sum_j f_2(b_j) p_j'
\end{equation}
holds for all initial states of the system, which yields the same
relation as in Eq.~\eqref{eqn:AVCP-case2-av-relation}.

Hence, combining the foregoing three sub-cases, we obtain the
\emph{generalized sum rule}:
\begin{multline} \label{eqn:AVCP-case-2-general-rule}
c(t_2) = f_1\left(a(t_1)\right) + f_2\left(b(t_1)\right) \\
\mapsto \ev{\cmatrix{C}}_{t_2} = \evb{f_1(\cmatrix{A})}_{t_1} +
                        \evb{f_2(\cmatrix{B})}_{t_1} \quad\forall \cvect{v}_0.
\end{multline}
In the special case where~$t=t_1 = t_2$, we obtain the \emph{sum
rule}:
\begin{equation} \label{eqn:AVCP-case-2-special-case-result}
c = f_1(a) + f_2(b) \mapsto  \cvect{C} =  f_1(\cvect{A}) +
                                                f_2(\cvect{B}).
\end{equation}
More generally, in the case of a classical experiment where measurements
of~$A^{(1)}, A^{(2)}, \dots, A^{(m-1)}$ are performed on a system at
time~$t_1$ and a measurement of~$A^{(m)}$ at time~$t_2$, with
values~$a^{(1)}, \dots, a^{(m)}$, respectively, the AVCP
implies the generalized operator rule:
\begin{multline} \label{eqn:AVCP-case2-av-relation-general-av}
a^{(m)}(t_2) = \sum_{l=1}^{m-1} f_l\bigl(a^{(l)}(t_1)\bigr)
\\
\mapsto \evb{\cmatrix{A}^{(m)}}_{t_2} = \sum_{l=1}^{m-1}
\evb{f_l(\cmatrix{A}^{(l)})}_{t_1} \quad \forall\cvect{v}_0.
\end{multline}
Taking the special case of simultaneous measurements~($t_1 =
t_2$), we obtain the operator rule
\begin{equation} \label{eqn:AVCP-case2-av-relation-general}
a^{(m)} = \sum_{l=1}^{m-1} f_l(a^{(l)}) \mapsto \cmatrix{A}^{(m)} =
\sum_{l=1}^{m-1} f_l(\cmatrix{A}^{(l)}).
\end{equation}

\subsubsection*{Case 3. $f(a,b)= f_1(a)f_2(b)$}
\label{sec:AVCP-case3}

We again consider three sub-cases. First, if measurements~\mment{A}
and~\mment{B} are represented by commuting operators and, if the
measurements are subsystem measurements and are performed on the
same subsystem, then, in the quantum experiment, they are performed
on the same copy of the system. Since~$f$ is, therefore, simple, the
AVCP applies, so that the relation
\begin{equation}
\begin{split}
    \sum_k c_k p_k''    &=  \sum_{i,j}  f_1(a_i)f_2(b_j)  p_i
                                                            p'_{j|i} \\
                        &=   \sum_i f_1(a_i) f_2(b_i) p_i
\end{split}
\end{equation}
holds for all initial states,~$\cvect{v}_0$, of the system.  Hence,
the generalized operator relation%
\begin{equation}
\ev{\cmatrix{C}}_{t_2}  =
\evb{f_1(\cmatrix{A})f_2(\cmatrix{B})}_{t_1}
\end{equation}
holds for all~$\cvect{v}_0$.  In the case where
measurements~\mment{A}, \mment{B}, and~\mment{C} are simultaneous,
we obtain the operator relation
\begin{equation}
    \cmatrix{C} = f_1(\cmatrix{A}) f_2(\cmatrix{B}).
\end{equation}

Second, if measurements~\mment{A} and~\mment{B} are represented by
commuting operators and are subsystem measurements performed on
different subsystems of a composite system, then they can be
performed on the same copy of the system in the quantum
experiment, in which case we obtain the same result as above.

Third, if measurements~\mment{A} and~\mment{B} are represented by
non-commuting operators, then the function~$f$ is not simple, and
the AVCP does not apply.

We can combine the three foregoing sub-cases to obtain the
 \emph{generalized product rule}:
\begin{multline} \label{eqn:AVCP-case-3-general-rule}
c(t_2) = f_1\bigl(a(t_1)\bigr)f_2\bigl(b(t_1)\bigr) \\
\mapsto
\ev{\cmatrix{C}}_{t_2} = \evb{f_1(\cmatrix{A})
                                f_2(\cmatrix{B})}_{t_1} \quad\forall
                                \cvect{v}_0\quad \text{if}~[\cmatrix{A},
                                \cmatrix{B}]=0.
\end{multline}
In the special case where~$t_1 = t_2$, we obtain the \emph{product
rule}:
\begin{equation} \label{eqn:AVCP-case-3-special-case-result}
c = f_1(a)f_2(b) \mapsto  \cmatrix{C} =  f_1(\cmatrix{A})
                                                f_2(\cmatrix{B})
                                                \quad \text{if}~[\cmatrix{A},
                                                            \cmatrix{B}]=0.
\end{equation}

\subsubsection*{Some comments on inconsistencies}
\label{sec:inconsistencies}

As mentioned in Sec.~\ref{sec:av-condition}, if the average-value
condition is weakened to allow  a measurement of~$AB$ to be
represented by an operator in the case where~$[\cmatrix{A},
\cmatrix{B}]\ne 0$, one is lead to a rule that is often stated,
namely
\begin{equation} \label{eqn:hermitization-rule}
f_1(a)f_2(b) \mapsto \frac{1}{2} \bigl(f_1(\cmatrix{A})
    f_2(\cmatrix{B}) + f_1(\cmatrix{A})f_2(\cmatrix{B}) \bigr)
\end{equation}
However, using this rule, one finds that inconsistencies quickly
arise.  For example, one can first apply this rule to find that
the operator representing a measurement of~$AB$ is
\begin{equation} \label{eqn:hermitized-form-AB2}
\widehat{AB} = (\cmatrix{A}\cmatrix{B} +
\cmatrix{B}\cmatrix{A})/2,
\end{equation}
where the notation~$\widehat{X}$ is used to denote the operator that
represents a measurement of~$X$. One can then apply the rule a
second time to find the operator that represents a measurement
of~$A^2B$. By treating this measurement as a measurement of~$A(AB)$,
or as a measurement of~$(A^2)B$, one obtains, respectively, either
\begin{equation}
\begin{split}
\widehat{A(AB)} &=\frac{1}{2} (\cmatrix{A}\widehat{AB} +
                                \widehat{AB}\cmatrix{A}) \\
                &=\frac{1}{4}
                \left(\cmatrix{A}(\cmatrix{A}\cmatrix{B} +
                \cmatrix{B}\cmatrix{A}) +
                (\cmatrix{A}\cmatrix{B} +
                \cmatrix{B}\cmatrix{A})\cmatrix{A} \right) \\
                &=\frac{1}{4} \left(\cmatrix{A}^2\cmatrix{B} +
                2\cmatrix{A}\cmatrix{B}\cmatrix{A} +
                \cmatrix{B}\cmatrix{A}^2 \right),
\end{split}
\end{equation}
or
\begin{equation}
\begin{split}
\widehat{(A^2)B} &=\frac{1}{2} (\widehat{A^2}\cmatrix{B} +
                                \cmatrix{B}\widehat{A^2}) \\
                &= \frac{1}{2} (\cmatrix{A}^2\cmatrix{B} +
                                \cmatrix{B}\cmatrix{A}^2),
\end{split}
\end{equation}
which are, in general, inequivalent.  Hence, the average-value
condition cannot be applied to non-simple functions of
observables, even in weakened form, without leading to
inconsistencies.

We also remark that, given the AVCP, it cannot consistently be
maintained that every classical measurement is
represented by a quantum measurement.  To see this, suppose that every classical measurement is represented by a quantum measurement.  Under this
supposition, the function and sum rules can be applied to a
measurement of~$(A+B)^2$, with~$[\cmatrix{A}, \cmatrix{B}]\ne 0$,
to derive Eq.~\eqref{eqn:hermitized-form-AB2} as follows. First,
defining~$d= a+b$, we use the sum rule to find~$\cmatrix{D} =
\cmatrix{A} + \cmatrix{B}$, and then use the function rule to find
that
\begin{equation}
\begin{split}
\widehat{D^2} &= \cmatrix{D}^2 \\
                &= \cmatrix{A}^2 + \cmatrix{A}\cmatrix{B} +
                    \cmatrix{B}\cmatrix{A} + \cmatrix{B}^2.
\end{split}
\end{equation}
Second, since, by assumption, every classical measurement is
represented by a quantum measurement, it follows that, in particular, 
a measurement of~$AB$ is represented
by a quantum measurement.  Therefore, we can use the sum rule directly to find a
measurement of~$D^2 = A^2 + 2AB + B^2$:
\begin{equation} \label{eqn:D-squared-2}
\begin{split}
\widehat{D^2} &= \widehat{A^2} + 2\widehat{AB} + \widehat{B^2} \\
                &= \cmatrix{A}^2 + 2\widehat{AB} + \cmatrix{B}^2.
\end{split}
\end{equation}
Equating these expressions for~$\widehat{D^2}$, we obtain
Eq.~\eqref{eqn:hermitized-form-AB2}, which, as we have seen, leads
to an inconsistency.  Hence, if the AVCP is accepted as valid, the original supposition must be false.  On the assumption that the AVCP is valid, this
inconsistency can only be avoided if we conclude that a
measurement of~$AB$
cannot be represented by a quantum measurement when~$[\cmatrix{A}, \cmatrix{B}]\ne 0$, in which case the
sum rule cannot be applied to obtain Eq.~\eqref{eqn:D-squared-2}.

In summary, given that measurements of~$A$ and~$B$ are represented
by quantum measurements~$\mment{A}$ and~$\mment{B}$, one can use
the AVCP to find quantum measurements that represent measurements
of~$f(A), f_1(A) + f_2(B)$ and, for~$[\cmatrix{A},
\cmatrix{B}]=0$, of~$f_1(A)f_2(B)$; and, more generally, one can
find quantum measurements that represent measurements of~$f(A,B)$
when~$f$ is simple. The AVCP also implies that a measurement
of~$AB$ cannot be represented by a quantum measurement
if~$[\cmatrix{A}, \cmatrix{B}]\ne 0$.

\subsection{Temporal Evolution}

In this section, we will use the AVCP, together with
the Temporal Evolution postulate~(see Paper~I), to derive the explicit form of the
temporal evolution operator for a system in a time-dependent
background.

As shown in Paper~I, temporal evolution of the system is represented by unitary
transformations. Specifically, over the course of the interval~$[t,
t+\Delta t]$, the state~$\cvect{v}(t)$ evolves as
\begin{equation}
    \cvect{v}(t+\Delta t) = \cmatrix{U}_t(\Delta t) \cvect{v}(t),
\end{equation}
where~$\cmatrix{U}_t(\Delta t)$ is the unitary matrix that
represents temporal evolution of the system during~$[t, t+\Delta
t]$.

Suppose now that the background of the system is time-independent
during this interval. Then we shall write~$\cmatrix{U}_t(\Delta
t)$ as~$\cmatrix{V}_t(\Delta t)$. Now, for~$0 \leq \Delta t_1 +
\Delta t_2 \leq \Delta t$, and~$\Delta t_1, \Delta t_2$  both
positive, we have
\begin{equation}
    \cmatrix{V}_t(\Delta t_1 + \Delta t_2)
        = \cmatrix{V}_{t+\Delta t_1}(\Delta t_2)
          \cmatrix{V}_t(\Delta t_1).
\end{equation}
But the time-independence of the background implies
that~$\cmatrix{V}_{t+\Delta t_1}(\Delta t_2) =
\cmatrix{V}_{t}(\Delta t_2)$.  Therefore,
\begin{equation}
\cmatrix{V}_t(\Delta t_1 + \Delta t_2)
        = \cmatrix{V}_{t}(\Delta t_2)
          \cmatrix{V}_t(\Delta t_1),
\end{equation}
which can be solved to yield
\begin{equation}
\cmatrix{V}_t(\Delta t_1) = \exp(-i\cmatrix{K}_t \Delta t_1),
\end{equation}
where~$\cmatrix{K}_t$ is a Hermitian matrix.

To determine the nature of~$\cmatrix{K}_t$, we proceed as follows.
In the classical model of a physical system in a time-independent
interval~$[t, t+\Delta t]$, the classical Hamiltonian for the
system is not explicitly dependent upon time during this interval.
On the assumption that the classical Hamiltonian is a simple
function of the observables of the system, and that the
measurements of these observables are represented by quantum
measurements~
\footnote{If the measurements of the observables of which the
classical Hamiltonian is a function are not represented by quantum
measurements, then it is not possible to write down the operator
corresponding to the classical Hamiltonian.  Similarly, if the
classical Hamiltonian is \emph{not} a simple function of the
observables of the system, then the form of the operator that
represents a measurement of energy cannot be determined by the
AVCP and, therefore, the explicit form of~$\cmatrix{U}_t(\Delta
t)$ cannot be obtained from the argument given in the text.
However, these assumptions do not appear to represent a
significant restriction in any fundamental cases of interest~(in
both non-relativistic and relativistic cases).},
it follows from the AVCP that the corresponding Hamiltonian
operator is also not explicitly dependent upon time during this
interval, so that, in particular,
\begin{equation} \label{eqn:Hamiltonian-constraint-1}
\cmatrix{H}_t = \cmatrix{H}_{t+\Delta t},
\end{equation}
where~$\cmatrix{H}_t$ denotes the Hamiltonian operator at
time~$t$. In addition, in the classical model, the total energy of
the system is constant during this interval for all states of the
system. Therefore, by the generalized function rule, the relation
\begin{equation} \label{eqn:Hamiltonian-constraint-2}
 \langle \cmatrix{H}_t \rangle_t  = \langle \cmatrix{H}_{t+\Delta t}
 \rangle_{t+\Delta t}
\end{equation}
holds for any state~$\cvect{v}$. Hence, from
Eqs.~\eqref{eqn:Hamiltonian-constraint-1}
and~\eqref{eqn:Hamiltonian-constraint-2}, it follows that
\begin{equation} \label{eqn:Hamiltonian-constraint-3}
 \langle \cmatrix{H}_t \rangle_t  = \langle \cmatrix{H}_t
 \rangle_{t+\Delta t}
\end{equation}
for all~$\cvect{v}$.  But
\begin{equation}
    \langle \cmatrix{H}_t \rangle_{t+\Delta t}
        =  \langle \cmatrix{H}_t \rangle_t +
        i \langle [\cmatrix{K}_t, \cmatrix{H}_t] \rangle_t \Delta t +
        \text{O}(\Delta t^2).
\end{equation}
Therefore, the commutator~$[\cmatrix{K}_t, \cmatrix{H}_t]=0$,
which implies that there exist~$N$ mutually orthogonal
eigenvectors,~$\cvect{v}_1, \dots, \cvect{v}_N$,
which~$\cmatrix{K}_t$ and $\cmatrix{H}_t$ share in common. In
particular, the state~$\cvect{v}_j$~$(j=1,2, \dots, N)$ is an
eigenvector of~$\cmatrix{H}_t$, with some eigenvalue~$E_j$.

Now, if the system is in an eigenstate,~$\cvect{v}_j$, with
eigenvalue~$k_j$, of~$\cmatrix{K}_t$, at time~$t$, the state
evolves as
\begin{equation} \label{eqn:v-K-evolution}
    \cvect{v}(t+\Delta t) = e^{-ik_j\Delta t}\,\cvect{v}(t).
\end{equation}
Therefore, during the interval~$[t, t+ \Delta t]$, this state
remains an eigenstate of~$\cmatrix{H}_t$, and is therefore a state
of constant energy,~$E_j$, during this interval.  In addition, since
evolution only affects the overall phase of the state, the
observable degrees of freedom of the state are time-independent
during this interval.  But, by the Temporal Evolution postulate, recalling from Paper~I 
that~$\phi_i = a\var_i + b$, the
state~$\cvect{v}(t)$, representing a system in a time-independent
background of definite energy,~$E_j$, whose observable degrees of
freedom are time-independent, evolves as
\begin{equation} \label{eqn:v-E-evolution}
  \cvect{v}(t+\Delta t) = e^{-iE_j\Delta t/\alpha'}\,\cvect{v}(t),
\end{equation}
where~$\alpha' = \alpha/a$.  
By comparison of Eqs.~\eqref{eqn:v-K-evolution}
and~\eqref{eqn:v-E-evolution}, we find that~$k_j = E_j/\alpha'$
holds for~$j=1, 2, \dots, N$, which implies that~$\cmatrix{K}_t =
\cmatrix{H}_t/\alpha'$.  Hence, in a time-independent background,
any state~$\cvect{v}$ evolves as
\begin{equation}
    \cvect{v}(t+\Delta t) =
        \exp\left(-i\cmatrix{H}_t\Delta t/\alpha' \right)
        \cvect{v}(t).
\end{equation}

In order to generalize to the case of temporal evolution in a
time-dependent background, we split the interval~$[t, t+\Delta t]$
into intervals of duration~$\epsilon$, approximate the evolution
during each of these intervals assuming that the background is
time-independent, and then take the limit as~$\epsilon \rightarrow
0$:
\begin{equation}
\begin{split}
    \cmatrix{U}_t(\Delta t) &= \cmatrix{U}_{t+\Delta t -
            \epsilon}(\epsilon) \dots \cmatrix{U}_{t+\epsilon}(\epsilon)
            \cmatrix{U}_{t}(\epsilon) \\
                            &=\lim_{\epsilon \rightarrow 0}
                            \bigl\{ \cmatrix{V}_{t+\Delta t -
            \epsilon}(\epsilon) \dots \cmatrix{V}_{t+\epsilon}(\epsilon)
            \cmatrix{V}_{t}(\epsilon) \bigr\},
\end{split}
\end{equation}
which, upon expansion, yields
\begin{equation}
    \cmatrix{U}_t(\Delta t) = I - \frac{i}{\alpha'}
                        \cmatrix{H}_t \Delta t + \text{O}(\Delta t^2),
\end{equation}
so that
\begin{equation}
    i\alpha'\frac{d\cvect{v}(t)}{dt}  =  \cmatrix{H}_t \cvect{v}(t).
\end{equation}
The value of the constant~$\alpha'$ will be determined at the end
of Sec.~\ref{sec:position-momentum-relations}.

\section{Representation of Measurements and Symmetry Transformations}
\label{sec:commutation-relations}

In this section, we shall use the AVCP to obtain~(a)~the commutation
relations~$[\cmatrix{x},\cmatrix{p}_x]=i\hbar$ and~$[\cmatrix{L}_x,
\cmatrix{L}_y]=i\hbar\cmatrix{L}_z$~(and cyclic permutations
thereof), and a restricted form of Dirac's Poisson bracket rule,
(b)~the explicit form of the displacement and rotation operators,
and (c)~the relation between the momentum and displacement
operators, and between the angular momentum and rotation operators.

From this point onwards, we shall take the infinite-dimensional
form of the abstract quantum formalism as a given.

\subsection{Position, momentum, and displacement operators}
\label{sec:position-momentum-relations}

We shall proceed in five steps.  First, by considering a particular
system~(a particle moving along the $x$-axis), we obtain the
commutation relationship,~$[\cmatrix{x}, \cmatrix{p}_x]=i\alpha'$.
Second, we obtain the explicit co-ordinate representation of
the displacement operator~$\cmatrix{D}_x = \frac{1}{i}
\frac{d}{dx}$.  Third, we obtain the commutation relations~$[\cmatrix{x}, \cmatrix{D}_x]=i$
and~$[\cmatrix{p}_x, \cmatrix{D}_x]=0$.  Fourth, we derive the 
relationship~$\cmatrix{D}_x = \cmatrix{p}_x/\alpha'$, and thereby obtain the co-ordinate
representation of~$\cmatrix{p}_x$. Fifth, we show that the
relations obtained are generally valid, and make the
identification~$\alpha' = \hbar$.

\subsubsection{The position--momentum commutation relationships}
\label{sec:position-momentum-relations-1}

Consider a massless particle moving in the $+x$-direction, where measurements
of the $x$-component of position, the $x$-component of momentum, and
the energy, are represented by the operators~$\cmatrix{x},
\cmatrix{p}_x, \cmatrix{H}$, respectively.

First, to determine the relationship of~$\cmatrix{H}$ to the
operators~$\cmatrix{x}$ and~$\cmatrix{p}_x$, we make use of the fact
that the relation~$H = cp_x$, where~$c$ is the speed of light, holds for all classical
states~$(x,p_x)$ of the system, so that, from the function rule, it
follows that~$\cvect{H} = c\cvect{p}_x$.

Next, to obtain a relation between~$\cmatrix{x}$
and~$\cmatrix{p}_x$, we make use of the fact that, in the quantum
model, the expected value of~$x$ at time~$t+\delta t$ can be
calculated in two separate ways. First, from the definition
of~$\ev{x}_t$, the relation
\begin{equation}
\begin{split}
 \ev{\cmatrix{x}}_{t+\delta t} &= \bigl\langle \cmatrix{U}_t^\dag(\delta t) \,\cmatrix{x}\, \cmatrix{U}_t(\delta t) \bigr\rangle_t       \\
                    &= \left\langle \left( 1+ \frac{i\cmatrix{H}\delta t}{\alpha'}\right)
                    \cmatrix{x}
                        \left( 1- \frac{i\cmatrix{H}\delta t}{\alpha'}\right) \right\rangle_t  + O(\delta t^2) \\
                    &= \ev{\cmatrix{x}}_t + \frac{i}{\alpha'} \delta t \ev{\cmatrix{Hx} - \cmatrix{xH}}_t      + O(\delta t^2)     \\
                    &= \ev{\cmatrix{x}}_t - \frac{ic}{ \alpha'} \delta t \evb{[\cmatrix{x},\cmatrix{p}_x]}_t + O(\delta
                    t^2)
\end{split}
\end{equation}
holds for all states,~$\cvect{v}$, of the system. Second, using the
generalized function rule, it follows from the classical
relation~$x(t+\delta t) = x(t) + c\delta t + O(\delta t^2)$ that the
relation
\begin{equation}
\ev{\cmatrix{x}}_{t+\delta t} = \ev{\cmatrix{x}}_t + c\delta t +
O(\delta t^2)
\end{equation}
holds for all~$\cvect{v}$.

Equating the above two expressions for~$\ev{x}_{t+\delta t}$, we
obtain
\begin{equation}
    \cvect{v}^\dagger \left[\cvect{x}, \cvect{p}_x\right] \cvect{v} = i\alpha'
\end{equation}
for all~$\cvect{v}$, which implies that
\begin{equation} \label{eqn:x-p-commutation-relation}
    \left[\cvect{x}, \cvect{p}_x \right] = i\alpha'.
\end{equation}
We note that, if one instead considers a particle of mass~$m$, moving non-relativistically in the~$x$ direction, so that the classical Hamiltonian given by~$H=p_x^2/2m + V(x)$, and~$x(t+\delta t) = x(t) + p_x(t)\delta t/m$, then the above computation yields the commutation relation~$\cmatrix{p}_x \left[\cmatrix{x}, \cmatrix{p}_x \right] +   \left[\cmatrix{x}, \cmatrix{p}_x \right]  \cmatrix{p}_x= 2i\alpha'\cmatrix{p}_x$, which can be solved to yield Eq.~\eqref{eqn:x-p-commutation-relation}.

Although a particular system has been used to obtain this
commutation relation, we shall later present an argument for its general
validity.

\subsubsection{Co-ordinate representation of the displacement
operator.}

Suppose that, in frame~$S$, the system is in state~$\psi(x)$.  The
probability density function over~$x'$ in the frame~$S'$, which is
displaced a distance~$-\epsilon$ along the $x$-axis, can be
calculated in two equivalent ways, according to whether the
transformation from frame~$S$ to~$S'$ is treated as a passive or
active transformation.  Accordingly, the probability density
function over~$x'$ can be obtained by performing measurements
of~$x'$ in frame~$S'$ upon the system in state~$\psi(x)$, or by
performing measurements of~$x$ in frame~$S$ upon the system in the
transformed state,~$\exp(-i\epsilon\cmatrix{D}_x)\psi(x)$, and
substituting~$x$ for~$x'$ in the resulting probability density
function over~$x$.

First, in frame~$S'$, let us calculate the probability density
function over~$x'$ directly.  In this frame, the
operator~$\cmatrix{x}'$ represents a measurement of~$x'$.  In the
classical model of the system, the relation
\begin{equation}
x' = x + \epsilon
\end{equation}
holds for all states~$(x,p_x)$ of the system.  Hence, by the
function rule, we obtain the operator relation
\begin{equation}
\cmatrix{x}' = \cmatrix{x} + \epsilon.
\end{equation}
Hence, an eigenstate of~$\cmatrix{x}$ with eigenvalue~$x$ is an
eigenstate of~$\cmatrix{x}'$ with eigenvalue~$x' = x + \epsilon$.
Therefore, if a measurement of~$x$ on a system in state~$\psi(x)$
yields values in the interval~$[x, x+ \Delta x]$ with
probability~$|\psi(x)|^2\Delta x$, then a measurement of~$x'$ on a
system in the same state yields values in the interval~$[x', x' +
\Delta x']$ with probability density
\begin{equation} \label{eqn:path1}
\Pr\left(x'|S', \psi(x) \right) =
\left|\psi(x'-\epsilon)\right|^2.
\end{equation}

Second, in frame~$S$, measurement~$x$ is performed on the system
in the transformed state~$\exp(-i\epsilon\cmatrix{D}_x) \psi(x)$,
so that the probability density function over~$x$ is
\begin{equation} \label{eqn:path2}
\Pr\left(x|S, \exp(-i\epsilon\cmatrix{D}_x)\psi(x) \right) =
                    \left|\exp(-i\epsilon\cmatrix{D}_x) \psi(x)\right|^2.
\end{equation}

The probability density functions over~$x'$ in
Eq.~\eqref{eqn:path1} and over~$x$ in Eq.~\eqref{eqn:path2}, must
agree under the correspondence~$x \leftrightarrow x'$.  Hence
\begin{equation}
\psi(x-\epsilon) = e^{i\phi(x)} \exp(-i\epsilon\cmatrix{D}_x)
                \psi(x),
\end{equation}
with~$\phi(x)$ being an arbitrary real-valued function of~$x$,
which is satisfied for any~$\epsilon$ if and only if~$\phi(x) = 0$
and
\begin{equation} \label{eqn:displacement-operator}
\cmatrix{D}_x = \frac{1}{i} \frac{d}{dx}.
\end{equation}

\subsubsection{The position-displacement and momentum-displacement commutation relations.}

In a classical model, the state of a particle subject to
measurements of $x$-position and the $x$-component of momentum, is
given by~$(x_0, p_{x0})$ in some frame of reference,~$S$. Consider
the following two experiments.

In the first experiment, measurements of the $x$~components of
position and momentum of the particle are made in a reference
frame,~$S'$, that is displaced by a distance~$\epsilon$ along the
$-x$~axis, resulting in the state~$(x',p_x')$ of the particle
relative to the co-ordinates of frame~$S'$. According to the
classical model,
\begin{equation} \label{eqn:classical-displacement-relations}
    \begin{aligned}
    x' &= x_0 + \epsilon  \\
    p_x' &= p_{x0}.
    \end{aligned}
\end{equation}
In the second experiment, the particle is displaced a
distance~$\epsilon$ in the $+x$-direction, and measurements of
position and momentum are then performed in frame~$S$, giving the
state,~$(x, p_x)$, of the particle in frame~$S$ as~$(x_0 +
\epsilon, p_{x0})$.

In classical physics, for all states of the particle, the
state~$(x', p_x')$, determined by measurements in frame~$S'$ upon
the undisplaced particle, is numerically identical to the state~$(x,
p_x)$, determined by measurements in frame~$S$ upon the displaced
particle.  That is,
\begin{equation} \label{eqn:classical-active-passive-equivalence}
    (x', p_x') = (x, p_x)
\end{equation}
for all states,~$(x_0, p_{x0})$, of the particle.

Now consider a quantum model of the particle subject to measurements
of~$x$ and~$p_x$, and let the state of the particle be given
by~$\cvect{v}_0$ in frame~$S$. Consider the first experiment. From
Eqs.~\eqref{eqn:classical-displacement-relations}, it follows from
the generalized function rule that, in the quantum model of the
particle, the relations
\begin{equation} \label{eqn:displacment-av-relations1}
    \begin{aligned}
    \ev{\cmatrix{x}'} &= \cvect{v}_0^\dagger  \cvect{x} \cvect{v}_0 + \epsilon \\
    \ev{\cmatrix{p}_x'} &= \cvect{v}_0^\dagger \cvect{p}_x \cvect{v}_0,
    \end{aligned}
\end{equation}
hold for all quantum states,~$\cvect{v}_0$, of the system.

In the second experiment, the displacement of the particle is a
continuous, symmetry transformation of the system, and therefore can
be represented by a unitary transformation of the
state,~$\cvect{v}_0$, and, in particular, by the
operator~$\exp\left( -i\epsilon \cvect{D}_x \right)$,
where~$\cvect{D}_x$ is a Hermitian operator.  To first order
in~$\epsilon$, measurements of~$x$ and~$p_x$ performed on this state
have expected values
\begin{equation} \label{eqn:displacment-av-relations2}
    \begin{aligned}
    \ev{\cmatrix{x}} &= \cvect{v}_0^\dagger (1+i\epsilon \cvect{D}_x) \cvect{x}
        (1-i\epsilon \cvect{D}_x) \cvect{v}_0 \\
    \ev{\cmatrix{p}_x} &= \cvect{v}_0^\dagger (1+i\epsilon \cvect{D}_x) \cvect{p}_x
        (1-i\epsilon \cvect{D}_x) \cvect{v}_0.
    \end{aligned}
\end{equation}

From Eq.~\eqref{eqn:classical-active-passive-equivalence}, by the
generalized function rule, the average values~$\ev{x'}$
and~$\ev{p_x'}$ of Eqs.~\eqref{eqn:displacment-av-relations1} are
respectively equal to the average values~$\ev{x}$ and~$\ev{p_x}$ of
Eqs.~\eqref{eqn:displacment-av-relations2} for all~$\cvect{v}_0$.
Hence, we obtain that the relations
\begin{equation} \label{eqn:displacment-av-relations3}
    \begin{aligned}
    \cvect{v}_0^\dagger (1+i\epsilon \cvect{D}_x) \cvect{x}
        (1-i\epsilon \cvect{D}_x) \cvect{v}_0  &= \cvect{v}_0^\dagger \cvect{x}
                                             \cvect{v}_0 + \epsilon \\
    \cvect{v}_0^\dagger (1+i\epsilon \cvect{D}_x) \cvect{p}_x
        (1-i\epsilon \cvect{D}_x) \cvect{v}_0 &= \cvect{v}_0^\dagger
                                    \cvect{p}_x \cvect{v}_0,
    \end{aligned}
\end{equation}
hold for all~$\cvect{v}_0$ to first order in~$\epsilon$, which
yield the commutation relations
\begin{equation} \label{eqn:displacment-commutation-relations}
    \begin{aligned}
    \left[ \cvect{x}, \cvect{D}_x \right] &= i \\
    \left[ \cvect{p}_x, \cvect{D}_x \right] &= 0
    \end{aligned}
\end{equation}

\subsubsection{The displacement-momentum relation and the co-ordinate representation of~$\cmatrix{p}_x$.}

From Eqs.~\eqref{eqn:x-p-commutation-relation}
and~\eqref{eqn:displacment-commutation-relations}, it follows that
\begin{equation}
\label{sec:displacement-minus-momentum-commutators}
    \begin{aligned}
    \left[\cvect{x}, \left(\cvect{D}_x - \cvect{p}_x/\alpha' \right)\right] &= 0 \\
    \left[\cvect{D}_x, \left(\cvect{D}_x - \cvect{p}_x/\alpha' \right) \right] &=
    0.
    \end{aligned}
\end{equation}

Now, in the co-ordinate representation, the
operators~$\cmatrix{x}$ and~$\cmatrix{D}_x$ are given by~$x$
and~$-i\alpha'\, d/dx$, respectively, and one can readily show
that~$\{x, -i\alpha'\, d/dx\}$ forms an irreducible
set~\footnote{See~\cite{Ballentine98}, Appendix~2.}. By Schur's
lemma~\footnote{See, for example, Ref.~\cite{Ballentine98},
Appendix~1.}, it therefore follows from
Eqs.~\eqref{sec:displacement-minus-momentum-commutators} that
\begin{equation} \label{eqn:displacement-momentum-relation1}
\cvect{D}_x = \frac{\cvect{p}_x}{\alpha'} + \gamma \cvect{I},
\end{equation}
where~$\gamma$ is real since the operator~$\left(\cvect{D}_x -
\cvect{p}_x/\alpha' \right)$ is Hermitian.  For a given
displacement,~$\epsilon$, the constant~$\gamma$ results in the
same overall shift of phase of any state,~$\cvect{v}$, of a
system, and therefore produces no physically observable effects on
the system.  Hence,~$\gamma$ can be set equal to zero without any
loss of generality, so that we obtain
\begin{equation} \label{eqn:displacement-momentum-relation2}
\cvect{D}_x = \frac{\cvect{p}_x}{\alpha'}.
\end{equation}
Analogous relationships for the displacement operator
corresponding to displacements in the~$y$ and~$z$ directions can
be obtained in a similar way.

Finally, from Eqs.~\eqref{eqn:displacement-operator}
and~\eqref{eqn:displacement-momentum-relation2}, we find
\begin{equation} \label{eqn:x-momentum-operator}
\cmatrix{p}_x = \frac{\alpha'}{i} \frac{d}{dx}.
\end{equation}

\subsubsection{Generality.} \label{sec:generality}

The representations of $x$- and~$p_x$-measurements have been
obtained above by considering, in the first step, a quantum model of
a particular physical system, namely a particle moving along the
$x$-axis.  In the general case of a particle moving in an arbitrary
direction, measurements of~$x, y, z$, and~$p_x, p_y, p_z$ are
subsystem measurements, and can therefore be represented in the
model of the composite system consisting of a particle, subject to
measurements chosen from a measurement set generated by a
measurement of~$\vect{r}=(x,y,z)$, by the operators~$x, y, z$
and~$-i\alpha' \,\partial/\partial x, -i\alpha' \,\partial/\partial y,
-i\alpha' \,\partial/\partial z$, respectively.

These representations of measurements of position and momentum are
also more generally valid for other systems, as we shall explain
below.

\medskip
\paragraph{State-determined measurements.}

Suppose that, in the classical framework, a measurement of~$A$ is
performed on a system, and the result is determined by the state of
the system alone.  That is, in particular, the result is
independent of the background of the system or of any
parameters~(such as charge or rest mass) that describe intrinsic
properties of the system.  We shall then say that this measurement
is a \emph{state-determined} measurement.  For example, the result
of a position measurement on a particle is determined by the state
of the particle, and is independent of whether or not the particle
is in an electromagnetic field and is independent of the mass or
charge of the particle.  In general, any measurement of an
observable that is a function only of the degrees of freedom of the
state of the system is a state-determined measurement.  In contrast,
a measurement of the total energy of a system is, in general,
dependent upon not only the state of the system, but also upon the
background of the system, and is therefore not a state-determined
measurement.

Now, consider two quantum models of two different physical
systems, system~1 and system~2, in different backgrounds, with
respect to the measurement set generated by
measurements~$\mment{A}_1$ and~$\mment{A}_2$, respectively,
where~$\mment{A}_1$ and~$\mment{A}_2$ represent a measurement
of~$A$ performed on the respective systems. Suppose, further, that
the two models have the same dimension. If the measurement of~$A$
is state-determined when performed on both systems~1 and~2, then,
by the AVCP, it follows that~$\ev{\cmatrix{A}_1} =
\ev{\cmatrix{A}_2}$ holds for all states,~$\cvect{v}$, where
operators~$\cmatrix{A}_1, \cmatrix{A}_2$ represent
measurements~$\mment{A}_1, \mment{A}_2$, respectively. It follows
at once that the operators~$\cmatrix{A}_1, \cmatrix{A}_2$ are
identical.

Hence, provided that two systems admit classical models with respect
to a measurement of~$A$ that is state-determined, and admit quantum
models of the same dimension with respect to
measurements~$\mment{A}_1$ and~$\mment{A}_2$, the operators that
represent~$\mment{A}_1$ and~$\mment{A}_2$ in the respective models
must be identical.

Therefore, if state-determined measurements of~$x$ and~$p_x$ are
performed on any system, then, in a quantum model of the system
subject to measurements in the measurement set containing quantum
measurements that represent measurements of~$x$ and~$p_x$, where
these measurements yield a continuum of possible results, their
representations are the same as those obtained above.  Therefore,
the commutation relations involving~$\cmatrix{x}, \cmatrix{p}_x$
and~$\cmatrix{D}_x$ are also generally valid.  Similar conclusions
clearly hold for measurements of~$y, z$ and~$p_y, p_z$.

Therefore, in the case of a particle where the interaction energy in
the Hamiltonian is obtained from a scalar potential that is
dependent on position only, in which case the measurements of
position and momentum are state-determined, the above
representations are valid. Below, we shall consider a physically
important case where the measurement of momentum is not
state-determined.

\medskip
\paragraph{Particle in a magnetic field.}

In the case of a charged particle in a magnetic field background
described in the Hamiltonian framework, the state of the particle
is~$(\vect{x}, \dot{\vect{x}})$, but the generalized co-ordinates
are taken to be~$(\vect{x}, \vect{p})$, where~$\vect{p} =
m\dot{\vect{x}} + e\vect{A}$, where~$m$ and~$e$ are the mass and
charge of the particle, respectively, and~$\vect{A}=(A_x, A_y, A_z)$
is the vector potential.  In this case,~$\vect{p}$ depends both upon
the state of the particle and the state of the background.
Therefore, a measurement of~$\vect{p}$ is not a state-determined
measurement, and the foregoing argument cannot be used to argue that
the operators representing the measurements of the components,~$p_x,
p_y, p_z$, of~$\vect{p}$ are those derived above. Instead, we reason
as follows.

First, for a particle with state~$(\vect{x}, \dot{\vect{x}})$ in a
magnetic field, in the argument of
Sec.~\ref{sec:position-momentum-relations-1}, the commutation
relation for the $x$-component of the motion in
Eq.~\eqref{eqn:x-p-commutation-relation} becomes
\begin{equation} \label{eqn:x-x-dot-commutator}
\bigl[\cmatrix{x}, m\dot{\cmatrix{x}} \bigr] = i\alpha',
\end{equation}
Then, from~$\vect{p} = m\dot{\vect{x}} + e\vect{A}$, the sum rule
gives
\begin{equation}
\cmatrix{p}_x = m \dot{\cmatrix{x}} + eA_x(\cmatrix{x},
\cmatrix{y}, \cmatrix{z}),
\end{equation}
which, together with Eq.~\eqref{eqn:x-x-dot-commutator}, implies
that
\begin{equation} \label{eqn:canonical-commutator-in-A-field}
\left[\cmatrix{x}, \cmatrix{p}_x \right] = i\alpha',
\end{equation}
as before.

Second, we note that, in the classical framework, the
momentum~$\vect{p}$ as defined above is invariant under
displacement of the reference frame.  Therefore,
Eqs.~\eqref{eqn:displacment-commutation-relations} remain
unchanged, and, using
Eq.~\eqref{eqn:canonical-commutator-in-A-field}, we
obtain~$\cmatrix{D}_x = \cmatrix{p}_x/\alpha'$.    Third, and
finally, the argument leading to the co-ordinate representation
of~$\cmatrix{D}_x$ remains unchanged since the argument only
involves measurements of position, which are state-determined
measurements. Therefore, the explicit representation
of~$\cmatrix{p}_x$ remains that given in
Eq.~\eqref{eqn:x-momentum-operator}, and similarly for the~$y$-
and~$z$-components of the motion.

\subsubsection{Identification of~$\alpha' = \hbar$.}

At this point, having obtained explicit representations for position
and momentum measurements, it is possible to use the operator rules
to write down the explicit Schroedinger equation for a structureless
electron in a hydrogen atom. By solution of the equation, and by
comparing the energy levels of the electron either with those found
in Bohr's model or with those found by experiment, one can establish
that the constant~$\alpha'$ is equal to~$\hbar$.

\subsubsection{Remark on applications.}

The correspondence rules derived above allow the quantum theoretic
modeling of a non-relativistic particle in an arbitrary classical
background consisting of gravitational and electromagnetic fields,
which leads to the non-relativistic Schroedinger equation. In the
case of a multi-particle system, the rules~(not discussed here)
for dealing with identical particles are, additionally, required.

In addition, the above rules allow the modeling of a photon without
consideration of polarization degrees of freedom~(leading to a
complex wave equation), a structureless relativistic
particle~(leading to the Klein-Gordon equation), and a relativistic
particle with internal degrees of freedom~(which, with the
appropriate auxiliary assumptions, leads to the Dirac equation).

\subsection{Angular momentum and rotation operators}

We shall proceed in three steps.  First, we shall obtain the
commutation relation~$[\cmatrix{L}_x, \cmatrix{L}_y] =
i\hbar\cmatrix{L}_z$, and cyclic permutations thereof, up to an
additive constant.  Second, we shall obtain the commutator relations
that hold between the rotation operators,~$\cmatrix{R}_x,
\cmatrix{R}_y, \cmatrix{R}_z$, and the angular momentum operators,
and shall then use these relations to determine the value of the
additive constant.  Third, we shall determine the relations that
hold between the rotation and angular momentum operators, and
indicate how the explicit representations of the angular momentum
operators and rotation operators can be determined.

\subsubsection{Components of Angular Momentum.}

Consider an experimental set-up where, in the classical model of the
set-up, measurements are performed upon a classical spin, with
magnetic moment~$\boldsymbol{\mu}$, which determine the values of
the rectilinear components of angular momentum of the system.
Suppose that the measurements of the components of angular momentum
along the $x$-, $y$-, and~$z$-directions, and the measurement of
energy, are represented by the operators~$\cmatrix{L}_x,
\cmatrix{L}_y, \cmatrix{L}_z$, and~$\cmatrix{H}$, respectively.

In particular, consider a set-up where a magnetic
field,~$\vect{B}$, is applied to the spin.  In the classical model
of this set-up, the energy associated with the interaction
is~$-\boldsymbol{\mu}\cdot\vect{B}$. Since~$\boldsymbol{\mu} = q
\vect{L}/2m$, where~$q$ and~$m$ are the charge and mass,
respectively, of the spin, and~$\vect{L}=(L_x, L_y, L_z)$ is its
angular momentum vector, the energy can be written
as~$-(q/2m)\,\vect{B}\cdot\vect{L}$. By the sum
rule~(Sec.~\ref{sec:AVCP-case2}), the quantum mechanical
Hamiltonian is given by
\begin{equation}
\cmatrix{H} = -\frac{q}{2m} (B_x\cmatrix{L}_x + B_y\cmatrix{L}_y +
B_z\cmatrix{L}_z),
\end{equation}
where~$B_x, B_y$ and~$B_z$ are the rectilinear components
of~$\vect{B}$.

The application of a magnetic field to a classical spin causes its
angular momentum vector,~$\vect{L}$, to rotate about the axis
along which the magnetic field is applied by an angle that is
proportional both to~$|\vect{B}|$ and to the duration for which
the field is applied.  Let the rotation matrix corresponding to a
rotation about axis~$a$ be denoted~$\rmatrix{R}_a(\theta)$,
where~$\theta$ is the angle of rotation. From the properties of
rotation matrices, it follows that
\begin{equation} \label{eqn:rotation-matrix-relations}
 \rmatrix{R}_x (\epsilon) \rmatrix{R}_y (\epsilon) -
    \rmatrix{R}_y (\epsilon) \rmatrix{R}_x (\epsilon) =
 \rmatrix{R}_z (\epsilon^2) - \rmatrix{I},
\end{equation}
where~$\epsilon$ is an infinitesimal angle,
and~$\rmatrix{R}_a(\epsilon)$ is an infinitesimal rotation which can
be implemented by application of a magnetic field~$\vect{B}$ along
the axis~\textit{a} for some time~$\delta t$.  Using this
relationship, it is possible to deduce the commutation relations
that hold between the quantum mechanical operators,~$\cmatrix{L}_x,
\cmatrix{L}_y, \cmatrix{L}_z$, in the following way.

The unitary evolution corresponding to the application of a
magnetic field~$\vect{B}$ to a spin for a time~$\delta t$ is
\begin{equation} \label{unitary-for-magnetic-field}
\cmatrix{U}(\delta t) = \exp \left(-\frac{i}{\hbar} \cmatrix{H}
\delta t \right).
\end{equation}
If magnetic fields of equal strength are applied for equal
times,~$\delta t$, along the $x, y,$ and~$z$-axes, respectively,
the corresponding unitary evolution is given to first order
in~$\delta t$, respectively, by
\begin{equation}
    \begin{aligned}
 \cmatrix{U}_1(\delta t) &= 1- \frac{i\epsilon}{\hbar} \cmatrix{L}_x  \\
 \cmatrix{U}_2(\delta t) &= 1- \frac{i\epsilon}{\hbar} \cmatrix{L}_y  \\
 \cmatrix{U}_3(\delta t) &= 1- \frac{i\epsilon}{\hbar} \cmatrix{L}_z.
    \end{aligned}
\end{equation}

Define~$\proj(\cvect{v})$ as the operation upon the quantum
state,~$\cvect{v}$, of a spin which returns a three-dimensional
vector,~$\ev{\vect{\cmatrix{L}}}$, with
components~$\ev{\cmatrix{L}_x}, \ev{\cmatrix{L}_y}$
and~$\ev{\cmatrix{L}_z}$.

If the application of a magnetic field,~$\vect{B} = B_z \vect{k}$,
say, to a classical spin causes a rotation of~$\vect{L}$ by
angle~$\theta$, then, by the generalized operator rule in
Eq.~\eqref{eqn:AVCP-case2-av-relation-general}, in the quantum model
of the spin, the application of the field rotates the
vector~$\ev{\vect{\cmatrix{L}}} = \proj(\cvect{v})$ by the
angle~$\theta$ about the z-axis. From
Eq.~\eqref{eqn:rotation-matrix-relations}, it therefore follows
that, for any~$\cvect{v}$,
\begin{multline}  \label{eqn:U-matrix-relations}
\proj(\cmatrix{U}_1(\epsilon) \cmatrix{U}_2(\epsilon) \cvect{v}) -
    \proj(\cmatrix{U}_2(\epsilon) \cmatrix{U}_1(\epsilon) \cvect{v} )
    = \\
    \proj(\cmatrix{U}_3(\epsilon^2) \cvect{v} ) - \proj (\cvect{v}).
\end{multline}
Using the definitions
\begin{align*}
    \cvect{v}_1 &= \cmatrix{U}_3(\epsilon^2) \cvect{v} =  \cvect{v} + \delta  \cvect{v}_1  \\
    \cvect{v}_2 &= \cmatrix{U}_1(\epsilon)\cmatrix{U}_2(\epsilon) \cvect{v}
                                         =  \cvect{v} + \delta  \cvect{v}_2     \\
    \cvect{v}_3 &= \cmatrix{U}_2(\epsilon)\cmatrix{U}_1(\epsilon) \cvect{v}
                                         =  \cvect{v} + \delta  \cvect{v}_3,     \\
\intertext{where}
    \delta  \cvect{v}_1 &= -\frac{i}{\hbar} \epsilon^2 \cmatrix{L}_z \cvect{v}    \\
    \delta  \cvect{v}_2 &=
        -\left[ -\frac{i}{\hbar} \epsilon (\cmatrix{L}_x + \cmatrix{L}_y)
                -\frac{1}{\hbar^2} \epsilon^2 \cmatrix{L}_x \cmatrix{L}_y
        \right] \cvect{v} \\
    \delta  \cvect{v}_3 &=
        -\left[ -\frac{i}{\hbar} \epsilon (\cmatrix{L}_x + \cmatrix{L}_y)
                -\frac{1}{\hbar^2} \epsilon^2 \cmatrix{L}_y \cmatrix{L}_z
        \right] \cvect{v},
\end{align*}
equation~\eqref{eqn:U-matrix-relations} becomes
\begin{equation}  \label{eqn:U-matrix-relations2}
\proj(\cvect{v} + \delta\cvect{v}_2) - \proj(\cvect{v} +
\delta\cvect{v}_3) = \proj(\cvect{v} + \delta\cvect{v}_1) -
\proj(\cvect{v}).
\end{equation}
Equating the $x$-components of this equation, we obtain
\begin{equation}
\cvect{v}^\dagger\cmatrix{L}_x (\delta \cvect{v}_2 - \delta
\cvect{v}_3) + (\delta \cvect{v}_2 - \delta \cvect{v}_3)^\dagger
\cmatrix{L}_x \cmatrix{v} = \cvect{v}^\dagger\cmatrix{L}_x \delta
\cvect{v}_1  + \delta \cvect{v}_1^\dagger \cmatrix{L}_x
\cmatrix{v},
\end{equation}
and, inserting the explicit forms of the~$\delta \cvect{v}_i$, we
obtain the commutation relation
\begin{subequations}
\begin{equation}  \label{eqn:an-operator-commutes-with-x-y-zA}
    \bigl[ \cmatrix{L}_x , \; \cmatrix{L}_z + \frac{i}{\hbar} [\cmatrix{L}_x,
        \cmatrix{L}_y ] \bigr]  =0.
\end{equation}
Equating the $y$- and $z$-components similarly, one obtains the
relations
\begin{align} \label{eqn:an-operator-commutes-with-x-y-z}
    \bigl[ \cmatrix{L}_y , \; \cmatrix{L}_z + \frac{i}{\hbar} [\cmatrix{L}_x,
        \cmatrix{L}_y ] \bigr] &=0 \\
    \bigl[ \cmatrix{L}_z , \; \cmatrix{L}_z + \frac{i}{\hbar} [\cmatrix{L}_x,
        \cmatrix{L}_y ] \bigr] &=0
        \label{eqn:an-operator-commutes-with-x-y-zC}
\end{align}
\end{subequations}
By inspection, the above commutation relations have the solution
\begin{subequations}
\begin{equation} \label{eqn:L-x-L-y-solution}
\bigl[ \cmatrix{L}_x, \cmatrix{L}_y \bigr] = i \hbar \cmatrix{L}_z
+
                                    i\gamma_1\cmatrix{I},
\end{equation}
where~$\gamma_1$ is real constant since the
operators~$i[\cmatrix{L}_x, \cmatrix{L}_y ]$ and~$\cmatrix{L}_z$ are
hermitian. We shall later show that this solution is, in fact, the
most general one.

The discussion leading to this result can be repeated to yield the
relations

\begin{align}
\bigl[ \cmatrix{L}_y, \cmatrix{L}_z \bigr] &= i \hbar \cmatrix{L}_x
                + i\gamma_2\cmatrix{I} \label{eqn:L-y-L-z-solution} \\
\bigl[ \cmatrix{L}_z, \cmatrix{L}_x \bigr] &= i\hbar\cmatrix{L}_y
        + i\gamma_3\cmatrix{I}.\label{eqn:L-z-L-x-solution}
\end{align}
\end{subequations}
In order to determine the values of $\gamma$-factors, we require
the commutation relations between the rotation
operators,~$\cmatrix{R}_x, \cmatrix{R}_y, \cmatrix{R}_z$,
and~$\cmatrix{L}_x, \cmatrix{L}_y, \cmatrix{L}_z$, which we shall
now derive.

\subsubsection{Rotation--angular momentum commutation relations.}

Let an infinitesimal clockwise rotation of a frame of reference by
angle~$\epsilon$ about the~$z$-axis be represented by unitary
transformation~$\exp(-i\epsilon \cmatrix{R}_z)$,
where~$\cmatrix{R}_z$ is Hermitian.

Now consider a set-up where measurements of~$L_x, L_y$, and~$L_z$
are performed on a system in the original and in the transformed
frame of reference.  In the classical model of this situation, the
results of the measurements performed in the original~(unprimed)
and transformed~(primed) frames are, to first order in~$\epsilon$,
related by
\begin{equation} \label{eqn:classical-rotation-connections}
\begin{pmatrix} L_x' \\ L_y' \\ L_z'\end{pmatrix} =
                                \begin{pmatrix}
                                    1 & -\epsilon & 0 \\
                                    \epsilon & 1 & 0 \\
                                    0 &   0 &   1
                                \end{pmatrix}
                                \begin{pmatrix}
                                    L_x \\ L_y \\ L_z
                                \end{pmatrix}.
\end{equation}

By the generalized operator rule in
Eq.~\eqref{eqn:AVCP-case2-av-relation-general}, it follows that, in
the quantum model of the situation, the relation
\begin{equation} \label{eqn:av-rotation-connections}
        \begin{pmatrix} \ev{\cmatrix{L}_x'} \\ \ev{\cmatrix{L}_y'} \\ \ev{\cmatrix{L}_z'}
        \end{pmatrix}
        =
        \begin{pmatrix}
            1 & -\epsilon & 0 \\
            \epsilon & 1 & 0 \\
            0 &   0 &   1
        \end{pmatrix}
        \begin{pmatrix}
        \ev{\cmatrix{L}_x} \\ \ev{\cmatrix{L}_y} \\ \ev{\cmatrix{L}_z}
        \end{pmatrix}
\end{equation}
holds for all states,~$\cvect{v}$, of the system.  Using the
relation
\begin{equation}
\ev{\cmatrix{L}_x'} = \cvect{v}^\dagger (1+ i\epsilon
\cmatrix{R}_z)\cmatrix{L}_x (1-  i\epsilon \cmatrix{R}_z) \cvect{v}
+ O(\epsilon^2),
\end{equation}
we find from Eq.~\eqref{eqn:av-rotation-connections} that
\begin{equation}
\ev{\cmatrix{L}_x} + i\epsilon \ev{[\cmatrix{R}_z, \cmatrix{L}_x]}
=
            \ev{\cmatrix{L}_x} - \epsilon \ev{\cmatrix{L}_y}
\end{equation}
for all~$\cvect{v}$, which implies that
\begin{subequations}
\begin{equation} \label{eqn:R-z-L-x-commutator}
\bigl[\cmatrix{R}_z, \cmatrix{L}_x \bigr] = i\cmatrix{L}_y.
\end{equation}
Proceeding similarly for~$\ev{\cmatrix{L}_y'}$
and~$\ev{\cmatrix{L}_z'}$, we obtain
\begin{align}
\left[\cmatrix{R}_z, \cmatrix{L}_y\right] &= -i\cmatrix{L}_x \label{eqn:R-z-L-y-commutator} \\
\left[\cmatrix{R}_z,\cmatrix{L}_z\right] &= 0.
                                \label{eqn:R-z-L-z-commutator}
\end{align}
\end{subequations}

The commutation relations for~$\cmatrix{R}_x$ and~$\cmatrix{R}_y$
can be obtained by parallel arguments.

\subsubsection{Angular momentum commutation relations.}

Left-multiplying Eq.~\eqref{eqn:L-z-L-x-solution}
by~$\cmatrix{R}_z$, and applying
Eqs.~\eqref{eqn:R-z-L-x-commutator}--\eqref{eqn:R-z-L-z-commutator},
we obtain
\begin{equation}
\bigl[\cmatrix{L}_z, \cmatrix{L}_x \bigr] \cmatrix{R}_z - i
\bigl[\cmatrix{L}_y, \cmatrix{L}_z \bigr] = \left(i\hbar
\cmatrix{L}_y + i\gamma_3 \cmatrix{I}\right)\cmatrix{R}_z + \hbar
\cmatrix{L}_x.
\end{equation}
Using the Eq.~\eqref{eqn:L-z-L-x-solution} right-multiplied
by~$\cmatrix{R}_z$, this implies that
\begin{subequations}
\begin{equation}
\bigl[\cmatrix{L}_y, \cmatrix{L}_z \bigr] = i\hbar \cmatrix{L}_x.
\end{equation}
Parallel arguments applied to Eqs.~\eqref{eqn:L-x-L-y-solution}
and~\eqref{eqn:L-y-L-z-solution} using the commutation relations
between~$\cmatrix{R}_x, \cmatrix{R}_y$ and~$\cmatrix{L}_x,
\cmatrix{L}_y, \cmatrix{L}_z$ yield
\begin{align}
\bigl[ \cmatrix{L}_z, \cmatrix{L}_x \bigr] &= i\hbar\cmatrix{L}_y \\
\bigl[ \cmatrix{L}_x, \cmatrix{L}_y \bigr] &= i \hbar
                                                    \cmatrix{L}_z.
\end{align}
\end{subequations}

\subsubsection{Explicit form of angular momentum operators}

From the classical relation~$L^2 = L_x^2 + L_y^2 + L_z^2$, it
follows from the sum rule that~$\cmatrix{L}^2 =\cmatrix{L}_x^2
+\cmatrix{L}_y^2+\cmatrix{L}_z^2$.  Using this relation and the
above commutation relations for~$\cmatrix{L}_x, \cmatrix{L}_y$
and~$\cmatrix{L}_z$, explicit representations of these operators for
finite~$N$ can be obtained and the irreducibility of the
representations of~$\cmatrix{L}_x, \cmatrix{L}_y, \cmatrix{L}_z$ can
be shown~\footnote{See~\cite{Group-Theory-in-Physics}~(Ch.~10,
Sec.~3), for example.}. Therefore, by Schur's lemma, the solution
given in Eq.~\eqref{eqn:L-x-L-y-solution} is the most general
solution of
Eqs.~\eqref{eqn:an-operator-commutes-with-x-y-zA}--\eqref{eqn:an-operator-commutes-with-x-y-zC},
and similarly for the solutions given in
Eqs.~\eqref{eqn:L-y-L-z-solution} and~\eqref{eqn:L-z-L-x-solution}.

Although the representations of~$\cmatrix{L}_x, \cmatrix{L}_y$
and~$\cmatrix{L}_z$ have been obtained by considering a particular
physical system, they are generally valid on account of the
argument given in Sec.~\ref{sec:generality}. Therefore the
commutation relations for~$\cmatrix{L}_x, \cmatrix{L}_y$
and~$\cmatrix{L}_z$ are generally valid.

\subsubsection{Rotation--angular momentum relations,
and explicit form of the rotation operators.}

Using the commutation relationships for~$\cmatrix{L}_x,
\cmatrix{L}_y$ and~$\cmatrix{L}_z$, it follows from
Eqs.~\eqref{eqn:R-z-L-x-commutator}--\eqref{eqn:R-z-L-z-commutator}
that~$(\hbar \cmatrix{R}_z - \cmatrix{L}_z)$ commutes
with~$\cmatrix{L}_x, \cmatrix{L}_y$, and~$\cmatrix{L}_z$.
Since~$\{\cmatrix{L}_x, \cmatrix{L}_y, \cmatrix{L}_z\}$ is an
irreducible set, it follows from Schur's lemma that
\begin{equation}
\hbar \cmatrix{R}_z - \cmatrix{L}_z = \gamma \cmatrix{I},
\end{equation}
where~$\gamma$ is a real-valued constant since~$\cmatrix{R}_z$
and~$\cmatrix{L}_z$ are Hermitian.  For any given~$\epsilon$, a
non-zero value of~$\gamma$ results in the same overall change of
phase for all states transformed
by~$\exp(-i\epsilon\cmatrix{R}_z)$, and so cannot give rise to
observable consequences. Hence,~$\gamma$ can be set to zero
without loss of generality. Therefore,~$\cmatrix{R}_z =
\cmatrix{L}_z/\hbar$, and, similarly,~$\cmatrix{R}_x =
\cmatrix{L}_x/\hbar$ and~$\cmatrix{R}_y = \cmatrix{L}_y/\hbar$.

Using the explicit representations of~$\cmatrix{L}_x,
\cmatrix{L}_y, \cmatrix{L}_z$ for any given dimension~$N$, the
explicit representation of the rotation operators follows at once
from these rotation--angular momentum relations. The explicit
co-ordinate representations of the rotation operators in the
infinite-dimensional case can also be determined by an argument
similar to that used earlier to determine the explicit form of the
displacement operators.

\subsection{Commutators and Poisson brackets}
\label{sec:PB-analogy}

In this section, we shall obtain a relation between the Poisson
Bracket,~$\{A, B\}$, and the commutator~$[\cmatrix{A},
\cmatrix{B}]$, where~$A$ and~$B$ are the classical observables of
a physical system describable in the classical Hamiltonian
framework, and~$\cmatrix{A}, \cmatrix{B}$ are the operators that
represent measurements of these observables. Dirac's Poisson
Bracket rule asserts the relation
\begin{equation} \label{eqn:Dirac-PB}
[\cmatrix{A}, \cmatrix{B}] = i\hbar \widehat{\{A, B\}},
\end{equation}
where~$\widehat{\{A,B\}}$ is the operator that represents a
measurement of~$\{A, B\}$.  Below, we shall derive this relation
using the AVCP in the case where~$B$ is the Hamiltonian.

Consider the Hamiltonian model of a system with state~$(q_1,
\dots, q_N; p_1, \dots, p_N)$ where~$N\ge 1$. The temporal rate of
change of the function~$F(q_1, \dots, q_N; p_1, \dots, p_N)$ is
given in terms of the Hamiltonian,~$H(q_1, \dots, q_N; p_1, \dots,
p_N)$, by
\begin{equation} \label{eqn:F-dot-PB}
\begin{split}
\dot{F} &=  \{F,H\} \\
        & = \sum_{i=1}^{N} \left\{ \frac{\partial F}{\partial q_i}
            \frac{\partial H}{\partial p_i} -  \frac{\partial H}{\partial q_i}
            \frac{\partial F}{\partial p_i} \right\}.
\end{split}
\end{equation}
Consider the quantum model of the system with state~$\cvect{v}$,
where the measurements of the~$q_i$ and the~$p_i$ are represented by
operators~$\cvect{q}_i$ and~$\cvect{p}_i$, respectively. If~$H$ is
simple, then, by the AVCP, a measurement of~$H$ can be represented
by the operator~$\cmatrix{H}$; otherwise, according to the AVCP, it
is not possible to describe the temporal evolution of the system in
the quantum model. If both of the functions~$\dot{F}$ and~$\{F, H\}$
are simple, then, by the AVCP, they are represented by the
operators~$\widehat{\dot{F}}$ and~$\widehat{\{F,H\}}$, respectively,
and from Eq.~\eqref{eqn:F-dot-PB}, by the generalized function rule,
the relation
\begin{equation} \label{eqn:PB1}
\evb{\widehat{\dot{F}}}_t =  \evb{\widehat{\{F,H\}}}_t,
\end{equation}
holds for all~$\cvect{v}$.

Now, in the classical model, the function~$\dot{F}$ is defined,
for all states, as
\begin{equation} \label{eqn:classical-F-dot}
\dot{F} =  \lim_{\Delta t \rightarrow 0} \left\{ \frac{F(t+\Delta
t) - F(t)}{\Delta t} \right\}.
\end{equation}
If~$F(t)$ and~$F(t+\Delta t)$ are both simple, then, according to
this definition,~$\dot{F}(t)$ is also simple, and, using the
generalized sum rule~(regarding the measurement of~$F(t+\Delta t)$
as the one being implemented in terms of measurements of~$F(t)$
and~$\dot{F}(t)$), we obtain the relation
\begin{equation} \label{eqn:PB2}
\evb{\widehat{\dot{F}}}_t =\lim_{\Delta t \rightarrow 0}
\frac{1}{\Delta t}
                    \left\{ \ev{\cmatrix{F}}_{t+\Delta t} - \ev{\cmatrix{F}}_t
                    \right\},
\end{equation}
which holds for all~$\cvect{v}$, with the operator~$\cmatrix{F}$
representing a measurement of~$F$.

If the functions~$\{F, H \}$ and~$F(t)$ are both simple, then,
since~$F(t+\Delta t) = F(t) + \{F, H \} \Delta t$, it follows
that~$F(t+\Delta t)$ is also simple. In that case, both
Eqs.~\eqref{eqn:PB1} and~\eqref{eqn:PB2} hold for all~$\cvect{v}$.
Equating these two expressions for~$\ev{\widehat{\dot{F}}}_t$, we
obtain the relation
\begin{equation} \label{eqn:PB-average-relation}
\begin{split}
    \evb{\widehat{\{F,H\}}}_t &=\lim_{\Delta t \rightarrow 0} \frac{1}{\Delta t}
                     \bigg\{ \left\langle \left( 1+ \frac{i}{\hbar} \cvect{H}\Delta t\right)
                        \cvect{F}
                     \left( 1- \frac{i}{\hbar} \cvect{H}\Delta t \right)
                     \right\rangle_t \\
                            &\quad\quad\quad\quad\quad\quad - \ev{\cmatrix{F}}_t  \bigg\} \\
                &= i\hbar^{-1} \left\langle [\cvect{H},\cvect{F}]
                \right\rangle_t,
\end{split}
\end{equation}
which holds for all~$\cvect{v}$, so that
\begin{equation} \label{eqn:PB-analogy}
    i\hbar \widehat{\{F,H\}}  =  [\cvect{F}, \cvect{H}].
\end{equation}
Hence, we obtain Eq.~\eqref{eqn:Dirac-PB} in the special case
where~$\cmatrix{B} = \cmatrix{H}$, subject to the condition that
the functions~$A, B$ and~$\{A,B\}$ are simple.  Using this relationship, we can readily evaluate useful commutation relationships.  For example, setting~$F = x$, and~$H = c p_x$, we find~$\{F,H\} = c$.  Hence, since the functions~$F, H$, and~$\{F,H\}$ are simple, Eq.~\eqref{eqn:PB-analogy} immediately gives~$[\cvect{x}, \cvect{p}_x] = i\hbar$.

If one or more of the functions~$F, H$, and~$\{F,H\}$ is not
simple, then Eq.~\eqref{eqn:PB-analogy} does not follow from the
above argument.
To take a specific example, suppose that, for a system with
state~$(x,p_x)$, where~$[\cmatrix{x}, \cmatrix{p}_x]=i\hbar$, we
choose~$F=x^3$ and~$H=\gamma p_x^3$, where~$\gamma$ is a constant.
We can then apply the function rule to obtain the corresponding
operators~$\cmatrix{F} =\cmatrix{x}^3$ and~$\cmatrix{H} =
\gamma\cmatrix{p}_x^3$,  and use these to find
\begin{equation} \label{eqn:pb-example-1}
[\cmatrix{F}, \cmatrix{H}] = 3i\gamma\hbar \left(\cmatrix{x}^2
\cmatrix{p}_x^2 + \cmatrix{x} \cmatrix{p}_x^2 \cmatrix{x} +
\cmatrix{p}_x^2 \cmatrix{x}^2\right).
\end{equation}
However, the function~$\{F,H\} = 9\gamma x^2p_x^2$ is not simple,
which implies that the AVCP cannot be used to write down an operator
which represents a measurement of~$\{F,H\}$.   If we were
nonetheless to apply the Hermitization rule in
Eq.~\eqref{eqn:hermitization-rule} to a measurement of~$\{F,H\}$~(in
spite of the inconsistencies to which we have shown this would lead)
we would obtain
\begin{equation} \label{eqn:pb-example-2}
i\hbar \widehat{\{F, H\}} = \frac{9i\gamma\hbar}{2}
\left(\cmatrix{x}^2\cmatrix{p}_x^2 + \cmatrix{p}_x^2 \cmatrix{x}^2
\right),
\end{equation}
but this differs from~$[\cmatrix{F}, \cmatrix{H}]$ by the
constant~$2\gamma\hbar^3$.  Since the expected value
of~$\gamma\hbar\cmatrix{x}^2\cmatrix{p}_x^2$ is itself of
order~$\gamma\hbar^3$, there is no guarantee that the difference
between Eqs.~\eqref{eqn:pb-example-1} and~\eqref{eqn:pb-example-2}
will be negligible.

\section{Discussion}

The derivation presented in this paper has shown that, using a
simple physical principle, it is possible to derive the
explicit form of the temporal evolution operator given the
postulates of Paper~I, and to derive the correspondence rules of
quantum theory in a systematic manner from appropriately chosen
relations known to hold in classical physics. The derivation
provides several physical insights into the correspondence rules.

The first insight is that the classical description of a
measurement~(such as `a measurement of~$x^2$') can be implemented by
more than one arrangement, and that, when
modeled in the quantum framework, these arrangements are, in
general, not equivalent.

Second, it is possible to impose a simple average-value condition
that must be satisfied by an operator that can be said to represent
an arrangement that implements a classical measurement.  This condition implies
that many arrangements cannot be represented by an operator, and
can therefore be eliminated from consideration. That is, one finds
that there are arrangements which, although acceptable representations of a measurement in the
classical framework, cannot be represented by operators in the
quantum framework without violating a very mild average-value
condition.

Third, in the case of an arrangement that satisfies the
average-value condition, the operator that represents the
arrangement is uniquely determined by the average-value condition
provided that the function,~$f$, that describes the arrangement,
is simple. One also finds that those arrangements that satisfy
the average-value condition are represented by the same
operator, so that it is possible to represent a measurement of~$f$
by a unique operator. If~$f$ is not simple, then it does not appear
to be possible to apply the average-value condition, even in a
weakened form, without inconsistencies arising.

Fourth,  we have found that the AVCP is incompatible with the
assumption that every classical measurement on a system
is represented by a quantum measurement in the quantum model of the
system.  For example, the AVCP implies that a measurement of~$AB$
does not have an operator representation if~$[\cmatrix{A},
\cmatrix{B}]\neq 0$.

The fifth insight rests on the fact that, rather surprisingly, the
AVCP enables correspondence rules of each of the four types
described in the Introduction~(operator rules, measurement commutation
relations, measurement--transformation commutation relations, and 
transformation operators) to be obtained in a uniform
manner. Consequently, one can see that the difference between
these types of rules depends simply upon whether the classical
relations that one is taking over into the quantum framework are
relations between measurements performed at the same time~(leading
to the operator rules), at different times~(leading to measurement commutation
relations), or in different frames of
reference~(leading to measurement--transformation commutation relations and to
the explicit forms of transformation operators). In short, from
the perspective provided by the derivation, the commutation
relation~$[\cmatrix{x}, \cmatrix{p}_x] = i\hbar$ is no more
mysterious in its origin than
the operator relation~$\cmatrix{H} = \cmatrix{p}_x^2/2m$.

Sixth, the derivation provides a clearer physical foundation to
many particular correspondence rules that are commonly used in
quantum theory. For example, the commutation
relation~$[\cmatrix{L}_x, \cmatrix{L}_y]=i\hbar \cmatrix{L}_z$ is
ordinarily derived in the infinite-dimensional quantum formalism
for a particle~(by transposing the classical
relation~$\cmatrix{L}_z = xp_y - yp_x$, and cyclic permutations
thereof, into the quantum framework using the operator rules), and
is then assumed, without further justification, to also hold in
the finite-dimensional case. Here, we have obtained this
commutation relation directly for finite- and infinite-dimensional
quantum systems, and have done so in a manner that makes clear its
connection with the properties of rotations. Similarly, a
restricted form of Dirac's Poisson bracket rule has been derived
in a systematic manner using the AVCP without making use of
abstract analogies.

Finally, we remark that, although the general notion of average-value
correspondence is already familiar in elementary quantum mechanics
through Ehrenfest's theorem, the possibility
that such a correspondence might serve as the basis for a
constructive principle that allows the correspondence rules of
quantum theory to be determined by appropriately-chosen classical
relations does not appear to have been widely
explored~\footnote{%
Refs.~\cite{vNeumann55, Groenewold-Principles-QM, Bohm51} mention
the general idea of average-value correspondence in their
discussion of the operator rules of quantum theory.  For example,
Groenewold~\cite{Groenewold-Principles-QM}~(Eqs.~(1.32)--(1.34))
remarks that the sum rule is equivalent to a condition on the
expectations of the respective operators, but the idea is not
formulated in a manner that is sufficiently systematic to derive
the operator rules, and no attempt is made to derive the any of
the other types of correspondence rule~(such as the measurement commutation relations)
using average-value correspondence.  Bohm~\cite{Bohm51} clearly
articulates the idea that average-value correspondence can be used
as a constraint on quantum theory, and uses it to determine
particular instances of the function rule~(Secs.~9.5--9.21) and to
determine the Hamiltonian operator that represents a
non-relativistic particle~(Secs.~9.24--9.26). However, the idea is
not
systematically formulated and applied beyond these special cases.}.  %
It has been shown here that it is possible to formulate the notion
of average-value correspondence in the form of an exact physical
principle which, roughly speaking, allows the logic of Ehrenfest's
argument to be reversed, enabling the often physically obscure
correspondence rules of quantum theory to be derived in a
systematic manner from familiar relations known to hold in
classical physics.

\begin{acknowledgments}

I am indebted to Steve Gull and Mike Payne for their support, and to Tetsuo Amaya and Matthew
Donald for discussions and invaluable comments.  I would like to thank the Cavendish Laboratory; Trinity College, Cambridge; Wolfson College, Cambridge; and Perimeter Institute for institutional and financial support.    Research at Perimeter Institute is supported in part by the Government of Canada through NSERC and by the Province of Ontario through MEDT.
\end{acknowledgments}

\end{document}